\documentclass[12pt]{article}

\pdfoutput=1

\usepackage[export]{adjustbox}

\usepackage{jheppub}
\usepackage{graphicx,epsfig,amsmath,amssymb}
\usepackage{subcaption}
\usepackage{multirow}
\usepackage{slashed}
\usepackage{cleveref}
\usepackage{enumitem}
\usepackage{calc}
\usepackage{placeins}
\usepackage[compat=1.1.0]{tikz-feynman}
\usetikzlibrary{positioning}
\usepackage{xcolor}

\usepackage{siunitx}
\usepackage{booktabs}
\usepackage{colortbl}
\usepackage{hyperref}

\def\as{\frac{\alpha_s}{2\pi}}

\def\be{\begin{equation}}
\def\ee{\end{equation}}
\def\bea{\begin{eqnarray}}
\def\eea{\end{eqnarray}}

\newcommand{\gosam}{\textsc{GoSam}{}}
\newcommand{\whizard}{\textsc{Whizard}{}}

\newcommand{\order}[1]{\ensuremath{\mathcal{O}\left(#1\right)}}

\title{Double Higgs Production in Vector Boson Fusion at NLO QCD in HEFT}
\author[a]{Jens Braun,}
\author[b]{Pia Bredt,}
\author[a]{Gudrun Heinrich,}
\author[a]{Marius H\"ofer}
 
\affiliation[a]{Institute for Theoretical Physics, Karlsruhe Institute
  of Technology (KIT), Wolfgang-Gaede-Str. 1, 76131 Karlsruhe, Germany}
\affiliation[b]{Department of Physics, University of Siegen,
  Walter-Flex-Straße 3, 57068 Siegen, Germany}

\emailAdd{j.braun@kit.edu}
\emailAdd{pia.bredt@uni-siegen.de}
\emailAdd{gudrun.heinrich@kit.edu}
\emailAdd{marius.hoefer@kit.edu}

\preprint{{\small KA-TP-03-2025, SI-HEP-2025-03,\\
    \hphantom{.}\hfill  P3H-25-010}}

\abstract{
  We present the next-to-leading order QCD corrections to Higgs boson pair production in vector boson fusion, including the leading operators in the framework of Higgs Effective Field Theory (HEFT).
  The corresponding calculation is based on an automated interface between the Monte Carlo event generator {\sc Whizard} and the one-loop amplitude generator {\sc GoSam}. The QCD corrections also include non-factorising diagrams and diagrams of Higgs-Strahlung type, thus going  beyond the structure function approach.
We find that some constellations of anomalous couplings, while being well within the current experimental constraints, can have a significant impact on the shape of typical observables for this process.}


\begin{document}

\maketitle

\section{Introduction}

The production of two Higgs bosons ranks among the most important processes at the Large Hadron Collider (LHC),  in particular in view of its High-Luminosity phase, as well as at prospective future colliders, because it allows us to shed more light on the trilinear Higgs-boson self-coupling, and thus on the Higgs potential.
While Higgs boson pair production through gluon fusion has the largest cross section among all di-Higgs production modes at the LHC, the production of a Higgs boson pair through vector boson fusion (VBF) has several appealing features:
experimentally motivated, as the two jets originating from the VBF process allow to tag on a distinct signature,
and theoretically motivated, as the coupling of Higgs bosons to vector bosons is intimately linked to unitarity and therefore small deviations in the couplings can have large effects.

The experimental collaborations already have put combined constraints on Higgs couplings to vector bosons and to itself, see Refs.~\cite{ATLAS:2024ish,CMS:2024awa,ATLAS:2024lsk,CMS-PAS-HIG-20-011} for recent results.

In the Standard Model (SM), QCD corrections to Higgs boson pair production through VBF have been calculated to very high order.
The next-to-leading order (NLO) QCD corrections were first calculated in Refs.~\cite{Figy:2008zd,Baglio:2012np,Frederix:2014hta} and are available in the codes {\tt VBFNLO}~\cite{Arnold:2008rz,Baglio:2024gyp} and {\tt MadGraph5\_aMC@NLO}~\cite{Alwall:2014hca}.
Differential next-to-next-to-leading order (NNLO) QCD corrections in the  in the so-called {\em structure function approximation}, where the process is treated as two factorising DIS-like topologies, have been calculated in Ref.~\cite{Ling:2014sne,Dreyer:2018rfu}, the total cross section is also available at next-to-next-to-next-to-leading order (N3LO)~\cite{Dreyer:2018qbw}.
In Ref.~\cite{Dreyer:2020xaj}, the NNLO QCD corrections are combined with NLO electroweak (EW) corrections.
The non-factorisable corrections appearing at NNLO, which can reach the 2\% level for large jet transverse momenta, have been calculated in the eikonal approximation in Ref.~\cite{Dreyer:2020urf} and have been implemented in the code {\tt proVBFHH}~\cite{proVBFHH}.

Opportunities to use the distinctive features of VBF Higgs pair production versus gluon fusion have been studied in Refs.~\cite{Dolan:2013rja,Dolan:2015zja}, 
a strategy to extract the $hhVV$ quartic coupling has been proposed in Ref.~\cite{Bishara:2016kjn}.

The importance of the VBF process for di-Higgs production within Effective Field Theories (EFT) to constrain anomalous Higgs couplings has been pointed out already some time ago~\cite{Ling:2017teo,Arganda:2018ftn,Araz:2020zyh,Kilian:2018bhs,Kilian:2021whd} and
has been studied under the aspect of distinguishing linear from non-linear realisations of the Higgs sector~\cite{Gomez-Ambrosio:2022qsi,Gomez-Ambrosio:2022why,Herrero:2022krh,Anisha:2024ljc,Anisha:2024ryj}.
The latter are also known under the name {\em Electroweak Chiral Lagrangian} or {\em Higgs Effective Field Theory\footnote{Which is not to be confused with the heavy top-quark limit of the SM, which is also sometimes referred to as HEFT.} (HEFT)}~\cite{Feruglio:1992wf,Bagger:1993zf,Koulovassilopoulos:1993pw,Burgess:1999ha,Wang:2006im,Grinstein:2007iv,Alonso:2012px,Buchalla:2012qq,Buchalla:2013rka}.


The publicly available Monte Carlo programs beyond LO for this process~\cite{Degrande:2020evl,Baglio:2012np,Baglio:2024gyp,Dreyer:2020xaj,proVBFHH} so far either did not 
 include anomalous couplings within an EFT framework, or did not include the full NLO QCD corrections without approximations.
With this paper we close this gap, presenting results that include the leading EFT operators within HEFT as well as the full NLO QCD corrections, including also the diagrams of Higgs-Strahlung type leading to the same $hhjj$ final state.
The code is based on a new version of the one-loop amplitude generator \gosam{}~\cite{Cullen:2011ac,GoSam:2014iqq,toappear} combined with the Monte Carlo event generator  \whizard~\cite{Moretti:2001zz,Kilian:2007gr} through the BLHA interface~\cite{Binoth:2010xt,Alioli:2013nda}, using a model file in Universal Feynman Output (UFO) format~\cite{Degrande:2011ua,Darme:2023jdn}.
The code also allows the combination with a parton shower, which is however not studied in this work.

The paper is organised as follows. In Section~\ref{sec:vbf_heft}, we discuss the process and the contributions of the leading HEFT operators. In Section \ref{sec:setup} we describe our implementation, before we discuss the results for the total and differential cross sections in Section~\ref{sec:results} and conclude.

\section{VBF in Higgs Effective Field Theory}
\label{sec:vbf_heft}

\subsection{Effective Lagrangian of Higgs Effective Field Theory}

Under the assumption that the energy scale of New Physics resides well above the electroweak scale of the SM, a parametrisation of beyond SM physics in terms of an EFT is well justified. Two such theories are commonly used in that context, the Standard Model Effective Field Theory (SMEFT)~\cite{Buchmuller:1985jz,Grzadkowski:2010es,Brivio:2017vri,Isidori:2023pyp} and the Higgs Effective Field Theory (HEFT)~\cite{Feruglio:1992wf,Bagger:1993zf,Koulovassilopoulos:1993pw,Burgess:1999ha,Wang:2006im,Grinstein:2007iv,Alonso:2012px,Buchalla:2012qq,Buchalla:2013rka}. The former organises the power counting of its effective operators in terms of their canonical mass dimension and is valid under the assumption that the otherwise unknown UV physics is weakly coupled~\cite{Arzt:1994gp,Craig:2019wmo,Buchalla:2022vjp}. HEFT on the other hand is suited also as an EFT description of UV theories with strong coupling in the Higgs sector, like for example composite Higgs models. As a consequence the Higgs field is treated independently from the Goldstone bosons of electroweak symmetry breaking, and the power counting is organised in the chiral dimension $d_\chi$, which is zero for bosons and one for each weak coupling, derivative and fermion bilinear~\cite{Buchalla:2013eza}:
\begin{align}
    d_\chi\left(X_\mu,S\right) &= 0\,, & d_\chi\left(\kappa,\partial,\bar{\psi}\psi\right) &= 1\,.
\end{align}
Here $X_\mu$ and $S$ represent generic gauge and scalar fields, respectively, and $\kappa$ can be any gauge or Yukawa coupling\footnote{The coefficients appearing in the functions $V(\eta)$, $F(\eta)$ and $\mathcal{M}(\eta)$ defined below are assumed to be radiatively generated and thus contain implicit factors of $\kappa$. See~\cite{Buchalla:2013eza} for details.}. For any given operator the chiral dimension directly corresponds to the operator's loop order $L$, $d_\chi=2L+2$. HEFT is particularly suited when one expects New Physics effects to show dominantly in anomalous Higgs couplings. The leading order (LO) HEFT Lagrangian collects all terms of chiral dimension $d_\chi=2$ and is given by~\cite{Buchalla:2013rka}
\begin{align}
    \mathcal{L}_2 =& -\frac{1}{4}G_{\mu\nu}^{a}G^{a\mu\nu} - \frac{1}{2}\left\langle W_{\mu\nu}W^{\mu\nu}\right\rangle - \frac{1}{4}B_{\mu\nu}B^{\mu\nu}\notag\\
        &+ \frac{v^2}{2}\partial_\mu\eta\,\partial^\mu\eta - V(\eta) + \frac{v^2}{4}\left\langle (D_\mu U)^\dagger (D^\mu U)\right\rangle F(\eta)\notag\\
        &+ \bar{\psi}i\slashed{D}\psi - \bar{\psi}\left(U\mathcal{M}(\eta)P_R + \mathcal{M}^\dagger(\eta) U^\dagger P_L\right)\psi\,,\label{eq:HEFT_L2}
\end{align}
with the $SU(3)_C$, $SU(2)_L$ and $U(1)_Y$ field strengths $G_{\mu\nu}^a$, $W_{\mu\nu}^\alpha$ and $B_{\mu\nu}$, and $\eta=h/v$, with $h$ being the Higgs singlet field and $v$ the scale of electroweak symmetry breaking. Angled brackets $\langle\dots\rangle$ denote a trace over $SU(2)_L$ group indices. For the sake of a compact notation the fermions are grouped into $\psi=(u,d,\nu,e)^T$, where each entry is a Dirac spinor and has to be understood as a vector in generation space. $P_R$ and $P_L$ are the usual right and left-handed projectors, such that $P_L\psi=(u_L,d_L,\nu_L,e_L)^T\equiv(Q_L,L_L)^T$ and $P_R\psi=(u_R,d_R,0,e_R)^T$, with $Q_L$, $L_L$ and $u_R$, $d_R$, $e_R$ the usual $SU(2)_L$ doublet and singlet fields, respectively. The fermions' covariant derivative is then given by
\begin{align}
    D_\mu\psi &= \left(\partial_\mu + ig_sG_\mu + igW_\mu P_L + ig'B_\mu(Y_LP_L+Y_RP_R)\right)\psi\,,
\end{align}
where $G_\mu=G_\mu^aT^a$, $W_\mu=W_\mu^\alpha t^\alpha$, with $T^a$ and $t^\alpha$ denoting the generators of $SU(3)_C$ and $SU(2)_L$, respectively. Their normalisation is given by $\mathrm{Tr}(T^aT^b)=\delta^{ab}/2$ and $\langle t^\alpha t^\beta\rangle=\delta^{\alpha\beta}/2$. The hypercharges are $Y_L=\mathrm{diag}(1/6,1/6,-1/2,-1/2)$ and $Y_R=\mathrm{diag}(2/3,-1/3,0,-1)$.

In HEFT the Higgs potential $V(\eta)$ is a Maclaurin series in $\eta$,
\begin{align}
    V(\eta) &= v^4\sum_{n=2}^\infty V_n\eta^n\,,
\end{align}
generating self-interactions with an arbitrary number of Higgs bosons. The Goldstone sector of the electroweak symmetry is realised by means of the object $U=\exp(2i\varphi/v)$, where $\varphi=\varphi^\alpha t^\alpha$ are the Goldstone bosons. The covariant derivative of $U$ induces the interaction between the electroweak gauge bosons and the Goldstones,
\begin{align}
    D_\mu U &= \partial_\mu U + igW_\mu U - ig'B_\mu Ut^3\,.
\end{align}
Here $g$ and $g'$ are the gauge couplings of $SU(2)_L$ and $U(1)_Y$, respectively. The so-called Flare function
\begin{align}
    F(\eta) &= 1 + \sum_{n=1}^\infty F_n\eta^n\,
\end{align}
generates interactions between electroweak gauge and Goldstone bosons and an arbitrary number of Higgs fields.

The kinetic term of quarks and leptons is as in the SM, while the extended Yukawa sector of HEFT is expressed in terms of the generalised, block diagonal mass matrix $\mathcal{M} = \mathrm{diag}\left(\mathcal{M}_u,\mathcal{M}_d,\mathcal{M}_\nu,\mathcal{M}_e\right)$, where the $\mathcal{M}_f(\eta)$ are functions of the Higgs field, again inducing interactions with an arbitrary number of Higgs bosons,
\begin{align}
    \mathcal{M}_f(\eta) &= m_f + \sum_{n=1}^\infty\mathcal{M}_{f,n}\eta^n\,.
\end{align}
We note that the LO HEFT Lagrangian~(\ref{eq:HEFT_L2}) does not generate anomalous couplings without at least one Higgs field. In particular, the gauge-fermion interactions remain SM-like, and there are no purely fermionic operators at this order. The exact SM Lagrangian can be recovered from~(\ref{eq:HEFT_L2}) by setting
\begin{align}
    V_2=V_3&=m_h^2/2v^2\,,   &   V_4&=m_h^2/8v^2\,,   &   V_{n\geq5}&=0\,,\notag\\
    F_1&=2\,,    &   F_2&=1\,,    &   F_{n\geq3}&=0\,,\notag\\
    \mathcal{M}_{f,1}&=m_f\,,  &   \mathcal{M}_{f,n\geq2}&=0\,.\notag
\end{align}

The LO HEFT Lagrangian~(\ref{eq:HEFT_L2}) is not renormalisable in the classical sense, that is, not all UV divergences appearing in loop amplitudes constructed from $\mathcal{L}_2$ can be absorbed by renormalising its parameters. However, the renormalisation can be carried out order by order in the loop-, or equivalently chiral-dimension expansion. At the one-loop order, for example, operators with chiral dimension $d_\chi=4$ are required to cancel all UV-divergences coming from $\mathcal{L}_2$. Those operators are collected in $\mathcal{L}_4$ and can be found in~\cite{Buchalla:2013rka,Sun:2022ssa,Graf:2022rco}\,. At $d_\chi=4$ we encounter modified fermion-gauge interactions and pure fermionic operators that were absent in $\mathcal{L}_2$. The complete one-loop renormalisation in HEFT has been studied in~\cite{Alonso:2017tdy,Buchalla:2017jlu,Buchalla:2020kdh}.

We remark that one can in fact construct another contribution at chiral dimension $d_\chi=2$, given by the operator
\begin{align}
    \mathcal{O}_{\beta_1} &\sim v^2\left\langle U^\dagger D_\mu Ut^3\right\rangle^2\left(1+\sum_{n=1}^\infty F_{\beta_1,n}\eta^n\right)\,,
\end{align}
with $\order{1}$ coefficients $F_{\beta_1,n}$. However, this operator violates the custodial symmetry and generates a tree level contribution to the electroweak $T$-parameter~\cite{Peskin:1991sw}. In the SM the custodial symmetry is an approximate symmetry which becomes exact in the limit of vanishing hypercharge and when the Yukawa couplings of isospin partners are equal, that is, $Y_u=Y_d$, and equivalently in the lepton sector. We then have $T=0$. Non-vanishing hypercharge and differences between Yukawa couplings of isospin partners affect the $T$-parameter, but only at the one-loop order and beyond. $T=0$ still holds at tree-level and deviations from this result due to radiative corrections are small. This is in line with the experimental bounds on the $T$-parameter~\cite{deBlas:2021wap}, which also indicate that the contribution from $\mathcal{O}_{\beta_1}$ must be suppressed. Therefore, it is customary to treat the operator as loop suppressed and assign it to $\mathcal{L}_4$.

\subsection{Effective Lagrangian for Higgs Pair Production in VBF}
In practical applications it is sufficient to consider only a subset of all effective operators appearing in the HEFT Lagrangian. Focusing on Higgs boson pair production in VBF, and assuming that the coupling of the Higgs boson to light quarks is not drastically enhanced compared to the SM, there are only three potentially anomalous interactions from $\mathcal{L}_2$, which can be parametrised as follows,
\begin{align}
    \mathcal{L}_\mathrm{eff} &\supset \left(2c_V\frac{h}{v}+c_{2V}\frac{h^2}{v^2}\right)\left(m_W^2W_\mu^+W^{-\mu}+\frac{1}{2}m_Z^2Z_\mu Z^\mu\right) +c_\lambda\frac{m_h^2}{2v}h^3\,.\label{eq:LO_eff}
\end{align}
In terms of the parameters introduced in the previous section we have $c_V=F_1/2$, $c_{2V}=F_2$ and $c_\lambda=2v^2V_3/m_h^2$. The normalisation of the $c_i$ is chosen such that they all equal to one in the SM case. The fact that $Z$ and $W^\pm$ share the same coupling is a consequence of the aforementioned conservation of the custodial symmetry.

 We do not include any operators from $\mathcal{L}_4$ in our analysis, which we motivate as follows. In principle it is true that, when performing an NLO calculation, also operators with chiral dimension $d_\chi=4$, i.e. one-loop order, have to be considered. Tree-level amplitudes with a single vertex insertion from $\mathcal{L}_4$ are of the same order as genuine one-loop amplitudes constructed from $\mathcal{L}_2$ vertices only. Here, however, we explicitly restrict ourselves to NLO in QCD. It follows that the one-loop corrections we encounter are $\order{g_s^2}$ relative to the Born diagrams, which are $\order{g_{ew}^4}$ for Higgs boson pair production in VBF\footnote{We use $g_{ew}$ to count general electroweak couplings, i.e.~both genuine gauge interactions and inverse powers of $v$ appearing in Higgs-vertices.}. It is therefore consistent to include only NLO HEFT operators contributing at the same power in the QCD coupling. Although the HEFT operators do not exhibit explicit powers in either the electroweak or strong coupling, one can still make a formal assignment based on the potential UV-origin of those operators. For example, each gauge field appearing in a given operator has to couple with the appropriate gauge coupling constant, which also holds for interactions with particles from the UV theory. Similarly, from power counting arguments it follows that each Higgs boson introduces a factor of $1/v\sim g_{ew}$. One can check that there are no operators from $\mathcal{L}_4$ with which we can construct tree topology diagrams matching the required configuration of external particles for Higgs pair production in VBF and the correct coupling powers for NLO QCD ($\order{g_{ew}^4g_s^2}$) at the same time.

We should comment in particular on the operators from the classes $X^2Uh$ and $XUhD^2$ in the nomenclature of~\cite{Buchalla:2013rka}, which couple a gauge field strength $X_{\mu\nu}$ to the electroweak Goldstones and an arbitrary number of Higgs bosons. From these operators we can get additional anomalous couplings between $W^\pm$, $Z$ and $h$ beyond the ones present in the LO effective Lagrangian~(\ref{eq:LO_eff}), but also local interactions between Higgs bosons and photons and gluons. With the latter we can construct diagrams with ordinary VBF-topology (see figure~\ref{fig:VBF_LO} in the next section) with the electroweak vector bosons replaced by gluons. Those diagrams are one-loop suppressed because the anomalous gluon-Higgs couplings are loop-generated. This implies that they are of $\order{g_{ew}^2g_s^4}$, since we can assign a power of $\order{g_{ew}g_s^2}$ ($\order{g_{ew}^2g_s^2}$) to the $ggh$ ($gghh$) couplings, motivated by the reasoning above. This does not match the coupling order of Higgs boson pair production in VBF and its NLO QCD corrections, so we do not include effective gluon-Higgs couplings in our analysis. Note that in the SM we can have a similar situation, with the gluons coupling to the Higgs boson via a top-quark loop: one-loop diagrams with the same external states as VBF can be constructed, but they are not considered as NLO QCD corrections to VBF.

A similar argument can be made for the $d_\chi=4$ local interactions between the electroweak gauge bosons, including the photon, and the Higgs boson. They are all at least $\order{g_{ew}^3}$, leading to one-loop suppressed diagrams of $\order{g_{ew}^6}$. Thus those contributions are at the same level as NLO EW corrections and therefore discarded.

We conclude that the effective Lagrangian~(\ref{eq:LO_eff}) is indeed sufficient to describe the leading anomalous EFT effects in Higgs boson pair production in VBF at NLO QCD. Hence there are no anomalous vertices with non-SM like Lorentz structures or field configurations and our setting coincides with that of the often used $\kappa$-framework~\cite{LHCHiggsCrossSectionWorkingGroup:2012nn,LHCHiggsCrossSectionWorkingGroup:2013rie} with its coupling modifiers. Note that the situation changes when the NLO EW corrections are considered and additional operators from $\mathcal{L}_4$ become relevant. Those will introduce new vertex structures not present in the SM, which can not be covered within the $\kappa$-framework. See, for example,~\cite{deBlas:2018tjm} for a more detailed comparison of HEFT with the $\kappa$-framework.

\subsection{Higgs Boson Pair Production in VBF at NLO QCD}

\begin{figure}
    \centering
    \begin{subfigure}[b]{0.5\textwidth}
        \centering
        \includegraphics[scale=1]{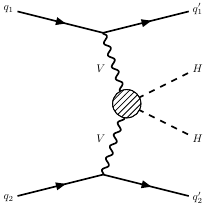}
        \caption{}\label{fig:VBF_LO_gen}
    \end{subfigure}%
    ~
    \begin{subfigure}[b]{0.5\textwidth}
        \centering
        $\includegraphics[scale=1,valign=c]{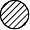} \in
        \left\{
        \includegraphics[scale=1,valign=c]{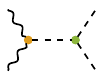},
        \includegraphics[scale=1,valign=c]{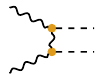},
        \includegraphics[scale=1,valign=c]{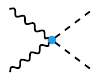}
        \right\}$
        \caption{}\label{fig:VBF_LO_sub}
    \end{subfigure}
    \caption{$pp\to hhjj$ at leading order. The shaded blob in the left diagram~(\subref{fig:VBF_LO_gen}) can represent any of the sub-diagrams~(\subref{fig:VBF_LO_sub}) on the right. We have massless quarks $q_i\in\{d,u,s,c,b\}$ and $V\in\{W^\pm,Z\}$. Crossings of $s$- and $u$-channel type are also allowed. See text for details.}\label{fig:VBF_LO}
\end{figure}

In this section we present all the contributions we considered for our calculation of Higgs boson pair production in VBF. Concretely, we define the process as electroweak Higgs boson pair production in proton-proton collisions in association with two light jets, i.e. $pp\to hhjj$ at $\order{\alpha_{ew}^4}$. We use $g_{ew}$ or $\alpha_{ew}=g_{ew}^2/4\pi$ to keep track of any kind of electroweak couplings in the Feynman diagrams. The NLO QCD corrections then comprise all contributions of $\order{\alpha_s}$ relative to the Born level. We take the quarks of the first two generations and the bottom quark to be massless and neglect any couplings between them and the Higgs boson, both in the SM case and when including EFT effects. We do not consider top quarks as external states, and assume a diagonal CKM-matrix. The Born process is then given by the diagrams in figure~\ref{fig:VBF_LO}.
Since in our setup the Higgs boson does not couple to light quarks, it only interacts with the internal vector bosons $V=W^\pm,Z$. A Higgs boson pair can then be produced in three ways, depicted in figure~\ref{fig:VBF_LO_sub}: Through a single $hVV$ coupling with a subsequent splitting $h^*\to hh$ via the trilinear self-coupling, through two separate $hVV$ couplings, or a single $hhVV$ coupling. As seen in the previous section, all three of the involved couplings can be anomalous, parametrised by $c_\lambda$, $c_{V}$ and $c_{2V}$. In figure~\ref{fig:VBF_LO_sub} they have been marked by green, orange and blue blobs, respectively.

We emphasise that, while the diagram in figure~\ref{fig:VBF_LO_gen} is what is usually understood as a VBF topology, our process definition mentioned above allows for any of the two quarks to be crossed into the initial state. This means that besides the depicted $t$-channel-like topology we also include $u$- and $s$-channel-like diagrams. The latter, $q_i\bar{q}_j\to V^* \to q_k\bar{q}_l$ with $V$ radiating two Higgs bosons, can be considered as a Higgs-strahlung topology instead. But since it matches the process definition in external states and coupling order, we will consider it in the following. Note, however, that diagrams of this kind are usually suppressed when VBF cuts are applied~\cite{Ciccolini:2007ec,Dreyer:2020xaj}.

\begin{figure}
    \centering
    \begin{subfigure}[b]{0.33\textwidth}
        \centering
        \includegraphics[scale=1]{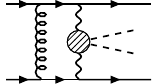}
        \caption{}\label{fig:VBF_NLO_B}
    \end{subfigure}%
~
    \begin{subfigure}[b]{0.33\textwidth}
        \centering
        \includegraphics[scale=1]{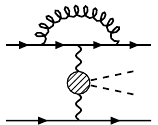}
        \caption{}\label{fig:VBF_NLO_A}
    \end{subfigure}%
~
    \begin{subfigure}[b]{0.33\textwidth}
        \centering
        \includegraphics[scale=1]{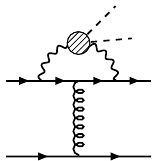}
        \caption{}\label{fig:VBF_NLO_C}
    \end{subfigure}%
    \caption{One-loop QCD corrections to $pp\to hhjj$. The shaded blob again represents any of the sub-diagrams in figure~\ref{fig:VBF_LO_sub}. $s$- and $u$-channel type crossings are allowed, too. See text for details.}\label{fig:VBF_NLO}
\end{figure}

At NLO QCD we have to consider both real and virtual (one-loop) corrections. The former can be constructed by attaching a single external gluon in any possible way to the quark lines in the Born process. The gluon can also be crossed into the initial state. For the virtual corrections on the other hand we have to consider all diagrams with an internal gluon added to the Born process. This gluon can be exchanged between the two separate quark lines as in figure~\ref{fig:VBF_NLO_B}, or be attached to a single quark line as in figure~\ref{fig:VBF_NLO_A}. The latter represent factorisable QCD corrections to the VBF process. In addition we can have diagrams as depicted in figure~\ref{fig:VBF_NLO_C}, where the two quark lines are connected through an internal gluon only and the two Higgs bosons are radiated from an electroweak-like loop. While this diagram does not have the usual VBF topology, it still has the correct external states and coupling order to be considered as part of the process.

Note that no additional anomalous couplings emerge at NLO QCD. Also, all anomalous couplings only appear in the sub-diagrams shown in figure~\ref{fig:VBF_LO_sub}, which do not receive any direct corrections at NLO QCD. The renormalisation of the amplitudes can be carried out as in the SM, no additional counterterms or modifications of counterterms are required.

\section{NLO Framework and Setup}
\label{sec:setup}
\subsection{Tools}
For the computation of cross sections and differential distributions we use an NLO framework which is composed of the event generator \whizard{}~\cite{Kilian:2007gr,Moretti:2001zz} and the one-loop-provider \gosam~\cite{Cullen:2011ac,GoSam:2014iqq} -- both operating for generic processes. 
For the analysis of the events and the generation of histograms we use {\sc Rivet}~\cite{Bierlich:2019rhm,Bierlich:2024vqo}.

\whizard{} is a multi-purpose Monte Carlo program providing all parts of event generation for hadron and lepton collider processes; from the theoretical model to showered and hadronised events. It comes along with the intrinsic tree-level matrix element generator \textsc{O'Mega} \cite{Moretti:2001zz} and the multi-channel phase space integrator {\sc Vamp}~\cite{Ohl:1998jn}, which is superseded by {\sc Vamp2}~\cite{Brass:2018xbv} with parallelisation capabilities based on the Message Passing Interface (MPI); this allows to run our simulation on multi-core architecture.
We apply \whizard's NLO automated framework \cite{ChokoufeNejad:2016qux, Stienemeier:2021cse} with real phase space construction and infrared subtraction following the FKS scheme \cite{Frixione:1995ms,Frixione:1997np}. In principle, parton shower matched NLO event samples can be produced by the automated {\sc powheg} matching~\cite{Nason:2004rx} implementation in \whizard~\cite{Stienemeier:2022wmy}, which however is not considered for this work.

The amplitudes at the tree and one-loop level are computed by \gosam{} and passed to \whizard{} via its interface based on the Binoth Les Houches Accord (BLHA) standard~\cite{Binoth:2010xt,Alioli:2013nda}.
For their construction \gosam{} applies algebraic methods to generate analytical expressions for Feynman diagrams with the help of {\sc qgraf}~\cite{Nogueira:1991ex} and {\sc form}~\cite{Vermaseren:2000nd,Kuipers:2012rf}. Subsequently, integral reduction is performed using the tool {\sc Ninja}~\cite{vanDeurzen:2013saa,Peraro:2014cba} and the master integrals are evaluated with {\sc OneLoop}~\cite{vanHameren:2010cp}.
The QCD UV renormalisation is done in an automated way in \gosam{}. The new version of \gosam~\cite{toappear} comes with improvements in the installation procedure and in speed; these features were also beneficial for the calculation presented here.

New code developments on the \whizard-\gosam{} interface allow the computation of NLO observables for user-defined hadron and lepton collider processes in the SM and beyond. For a consistent setup of both tools in view of New Physics models or Effective Field Theories, like HEFT or SMEFT, external model files are a key ingredient. All operators and vertices resulting from such theories can be summarised in files written in UFO format~\cite{Degrande:2011ua,Darme:2023jdn}, and thus are accessible by both tools via their separate UFO interfaces. 
The \whizard-\gosam{} interface has been extended to be able to cope with processes based on large classes of UFO model files.

Calculations with our setup can be done out-of-the-box providing a \whizard{} run file written in \texttt{Sindarin}, the steering language of this event generator, and an UFO model file in case the application of a theory beyond the SM is desired.
%
We provide a repository\footnote{\url{https://github.com/Jens-Braun/VBF_HH_HEFT}} 
that contains the UFO model file and an example run file for the process considered here.
Differential distributions at LO and NLO in the form of histograms, ultimately, can be produced by generating event samples with \whizard{} which are piped into a \textsc{Rivet} analysis in the next step.

\subsection{Validation}

We validated our \whizard-\gosam{} NLO framework by comparing several processes in the SM and in an EFT setup to the literature or to existing benchmarks. 
The setup using UFO model files has been extensively tested by comparing the results based on a SM UFO file to the case where the internal, well-validated, SM files have been used. As an example for an EFT setup, we carried out a comparison to the results of Ref.~\cite{Maltoni:2016yxb} for the process $pp\to t\bar{t}H$ in SMEFT; the details of this comparison can be found in~\cite{masterthesisMarijn}.

For the process considered here, we compared our setup at LO in the SM with \texttt{MadGraph5\_aMC@NLO} and \whizard+OpenLoops, showing very good agreement. We also compared to \texttt{VBFNLO} and found agreement except in the tail of the $p_\perp^h$ distribution,
which we attribute to the fact that 
\texttt{VBFNLO} uses the structure function approach.  
At NLO in the SM, a quantitative comparison at cross section level is difficult because the public tools that were available make approximations: if not the VBF approximation, then at least with regards to the treatment of the pentagon- and hexagon-integrals.
Therefore we used OpenLoops~\cite{Cascioli:2011va,Buccioni:2019sur} to check our setup at amplitude level, and in combination with \whizard{} at the cross section level, yielding again good agreement. 
In particular, comparing \gosam{} and OpenLoops at amplitude level, we have evaluated $10^6$ phase-space points for each, the Born amplitude squared, the real amplitude squared and the one-loop amplitude interfered with the Born amplitude. 
We found relative differences of ${\cal O}(10^{-15})$ for the Born- and real radiation amplitudes and of ${\cal O}(10^{-8})$ for the finite part of the virtual amplitude.
 More details about the validation can be found in Ref.~\cite{masterthesisJens}.

\section{Phenomenological Results}
\label{sec:results}

\subsection{Input Parameters}
For the numerical evaluation of the process, we consider proton-proton collisions at the LHC with a centre-of-mass energy of $\sqrt{s} = \qty{13.6}{\TeV}$. We use the \texttt{PDF4LHC21\_mc} parton distribution function (PDF) set \cite{PDF4LHCWorkingGroup:2022cjn} obtained through LHAPDF6 \cite{Buckley:2014ana} with the associated $\alpha_s(\mu)$. As the set of electroweak input parameters we use $\{\alpha, G_F, m_Z\}$, with the respective values of $\alpha^{-1}(m_Z) = 127.9$ and $G_F = \SI{1.16637e-5}{\GeV^{-2}}$. The masses and widths of the contributing particles, including the input parameter $m_Z$, are set to 
\begin{equation}
    \begin{gathered}
    m_Z = \SI{91.1876}{\GeV}, \quad \Gamma_Z = \SI{2.4952}{\GeV}, \\ 
    m_W(\alpha, m_Z, G_F) = \SI{79.8244}{\GeV}, \quad \Gamma_W = \SI{2.085}{\GeV}, \\
    m_h = \SI{125.0}{\GeV}, \quad \Gamma_h = \SI{4.07}{\MeV}.
    \end{gathered} \label{eq:masses_widths}
\end{equation} 
 Notably, in this scheme the above value of $m_W$ is not used as input but calculated from the input parameters by means of SM tree-level relations, which are not altered by any EFT contribution at the order we are considering. We use a diagonal CKM-matrix and exclude top quarks from the set of possible external states.

Scale uncertainties are included by means of a three-point scale variation of the renormalisation and factorisation scales, $\mu_R = \mu_F = \xi \mu_0$ with $\xi \in \{1/2 , 1, 2\}$ and the central scale being chosen as in \cite{Dreyer:2018rfu},
\begin{equation}
    \mu_0 = \sqrt{\frac{m_h}{2} \sqrt{\frac{m_h^2}{4} + p_{\bot, hh}^2}}\;. \label{eq:scale}
\end{equation}
In~\cite{Dreyer:2018rfu} it is argued that this scale approximates well the square root of the product of the virtualities of the two vector bosons.

Jets are clustered with the anti-$k_T$ algorithm \cite{Cacciari:2008gp} implemented in the \texttt{FastJet} package \cite{Cacciari:2011ma} with a radius parameter of $R = 0.4$. We require the presence of at least two jets satisfying 
\begin{equation}
    p_{\bot,j} > \qty{20}{\GeV} \quad \text{and} \quad |y_j| < 4.5.
\end{equation}
We apply some additional cuts on the invariant mass and pseudo-rapidity separation of the two hardest jets $j_1$ and $j_2$, referred to as \emph{tagging jets}, \begin{equation}
    m_{j_1j_2} > \qty{600}{\GeV} \quad \text{and} \quad |\eta_{j_1} - \eta_{j_2}| > 4.0.
\end{equation}

\subsection{Total Cross Section}
\label{sec:tot_x_sec}

\begin{table}
    \caption{LO and NLO cross sections at a centre-of-mass energy of $\sqrt{s} = \qty{13.6}{\TeV}$ for various values of the anomalous couplings. The asymmetrical errors are calculated with a three-point scale variation, the symmetrical error is the MC integration error. For the ratios in the last two columns, only the MC error is given.}
    \centering
    \scalebox{0.9}{
    \renewcommand{\arraystretch}{1.2}
    \begin{tabular}{c | c c c | c c c c} \toprule
        No. & $c_\lambda$ & $c_V$ & $c_{2V}$ & $\sigma_\mathrm{NLO}$ in fb & $\sigma_\mathrm{LO}$ in fb & $\sigma_\mathrm{NLO}/\sigma_\mathrm{LO}$ & $\sigma_\mathrm{NLO}/\sigma_\mathrm{NLO}^\mathrm{SM}$ \\ \midrule
                             SM & 1.0  & 1.0  & 1.0 & $0.752^{+0.000}_{-0.012} \pm 0.005$ & $0.832^{+0.079}_{-0.074} \pm 0.002$ & 0.904(6) & 1.00(1) \\
        \rowcolor[gray]{0.9}  1 & -1.0 & 0.9  & 1.5 & $2.11^{+0.00}_{-0.07} \pm 0.01$     & $2.321^{+0.297}_{-0.250} \pm 0.003$ & 0.910(4) & 2.81(2) \\
                              2 & -1.0 & 1.05 & 1.3 & $3.40^{+0.03}_{-0.10} \pm 0.02$     & $3.753^{+0.310}_{-0.286} \pm 0.007$ & 0.905(5) & 4.51(4) \\
        \rowcolor[gray]{0.9}  3 & 0.0  & 1.0  & 1.0 & $1.94^{+0.01}_{-0.06} \pm 0.01$     & $2.139^{+0.189}_{-0.173} \pm 0.004$ & 0.907(7) & 2.58(3) \\
                              4 & 1.0  & 0.9  & 1.0 & $0.380^{+0.000}_{-0.014} \pm 0.003$ & $0.412^{+0.051}_{-0.038} \pm 0.002$ & 0.920(7) & 0.505(5) \\
        \rowcolor[gray]{0.9}  5 & 1.0  & 1.0  & 0.5 & $4.38^{+0.03}_{-0.07} \pm 0.02$     & $4.850^{+0.549}_{-0.445} \pm 0.007$ & 0.902(4) & 5.82(5) \\
                              6 & 1.0  & 1.0  & 1.5 & $1.52^{+0.00}_{-0.06} \pm 0.01$     & $1.675^{+0.215}_{-0.177} \pm 0.002$ & 0.910(4) & 2.03(2) \\
        \rowcolor[gray]{0.9}  7 & 2.0  & 0.9  & 1.4 & $3.86^{+0.02}_{-0.09} \pm 0.01$     & $4.276^{+0.497}_{-0.402} \pm 0.004$ & 0.902(3) & 5.13(4) \\
                              8 & 2.0  & 1.0  & 1.0 & $0.627^{+0.000}_{-0.035} \pm 0.005$ & $0.690^{+0.069}_{-0.063} \pm 0.002$ & 0.908(8) & 0.833(8) \\
        \rowcolor[gray]{0.9}  9 & 3.0  & 1.1  & 0.5 & $4.40^{+0.00}_{-0.13} \pm 0.02$     & $4.793^{+0.608}_{-0.488} \pm 0.006$ & 0.918(4) & 5.85(5) \\
                             10 & 4.0  & 0.95 & 0.5 & $2.41^{+0.01}_{-0.05} \pm 0.01$     & $2.653^{+0.269}_{-0.227} \pm 0.005$ & 0.907(5) & 3.12(3) \\
        \rowcolor[gray]{0.9} 11 & 6.0  & 1.1  & 1.0 & $9.82^{+0.00}_{-0.25} \pm 0.07$     & $10.95^{+1.03}_{-0.74} \pm 0.02$    & 0.896(7) & 13.1(1) \\ \bottomrule
    \end{tabular}
    \renewcommand{\arraystretch}{1}
    \label{tab:xsecs}
    }
\end{table}
First, in table \ref{tab:xsecs}, we present the total cross sections at LO and at NLO for several selected benchmark points in $(c_\lambda, c_V, c_{2V})$-space. The ranges in which the anomalous couplings are varied are oriented at the constraints from refs.~\cite{ATLAS:2024ish,CMS:2024awa}. Concretely, we use
\begin{equation}
c_\lambda \in [-1,6]\; , \; c_V\in [0.9, 1.1]\; , \; c_{2V}\in [0.5, 1.5]\;.
\end{equation}

We can immediately observe a significant dependence of the total cross section on the anomalous couplings, reaching an enhancement of more than an order of magnitude for benchmark point 11 or a decrease to less than half for benchmark point~4. In contrast to this, the K-factor depends only very mildly on the coupling values, varying by at most a few per cent. For the scale uncertainties, we can observe a clear reduction from LO to NLO.

To achieve a continuous description of the total cross section with respect to the anomalous couplings, we parametrise it as a polynomial in all possible combinations of anomalous couplings that can occur in the squared matrix element,
\begin{equation}
    \frac{\sigma}{\sigma^\mathrm{SM}} = A_0\ c_\lambda^2 c_V^2 + A_1\ c_V^4 + A_2\ c_{2V}^2 + A_3\ c_\lambda c_V^3 + A_4\ c_\lambda c_V c_{2V} + A_5\ c_V^2 c_{2V}. \label{eq:parameterisation}
\end{equation}
Note that, since anomalous couplings only appear in the sub-diagrams shown in figure~\ref{fig:VBF_LO_sub}, the functional form of this parametrisation is the same at LO and NLO. We obtain values for the coefficients $A_i$ by fitting the parametrisation \eqref{eq:parameterisation} to the cross sections of the benchmark points in table \ref{tab:xsecs} using the Python package \texttt{iminuit} \cite{iminuit,James:1975dr}. We repeat this procedure for $\mu_R = \mu_F = \mu_0/2$ and $\mu_R = \mu_F = 2 \mu_0$, resulting in the coefficients shown in table \ref{tab:coefficients}.

\begin{table}
    \caption{Fit results for the coefficients of equation \eqref{eq:parameterisation} with the central scale given by equation \eqref{eq:scale}. The uncertainties are those obtained in the fitting procedure.}
    \centering
    \scalebox{0.9}{
    \renewcommand{\arraystretch}{1.2}
    \begin{tabular}{c | c c c} \toprule
         Parameter & $\mu_F = \mu_r = \mu_0/2$ & $\mu_F = \mu_r = \mu_0$ & $\mu_F = \mu_r = 2 \mu_0$ \\ \midrule
                             $A_0$ & \num{0.687(7)}  & \num{0.693(5)} & \num{0.699(5)} \\
         \rowcolor[gray]{0.9}$A_1$ & \num{21.8(2)}   & \num{21.9(2)}   & \num{22.0(2)} \\
                             $A_2$ & \num{11.6(1)}   & \num{11.7(1)}   & \num{11.7(1)} \\
         \rowcolor[gray]{0.9}$A_3$ & \num{-6.04(6)}  & \num{-6.07(4)} & \num{-6.13(5)} \\
                             $A_4$ & \num{3.80(4)}   & \num{3.82(3)}  & \num{3.86(3)} \\
         \rowcolor[gray]{0.9}$A_5$ & \num{-30.9(3)}  & \num{-31.0(2)} & \num{-31.2(2)} \\ \bottomrule
    \end{tabular}
    \renewcommand{\arraystretch}{1}
    }
    \label{tab:coefficients}
\end{table}

\begin{figure}
    \begin{subfigure}[l]{0.49\textwidth}
        \includegraphics[width=\textwidth]{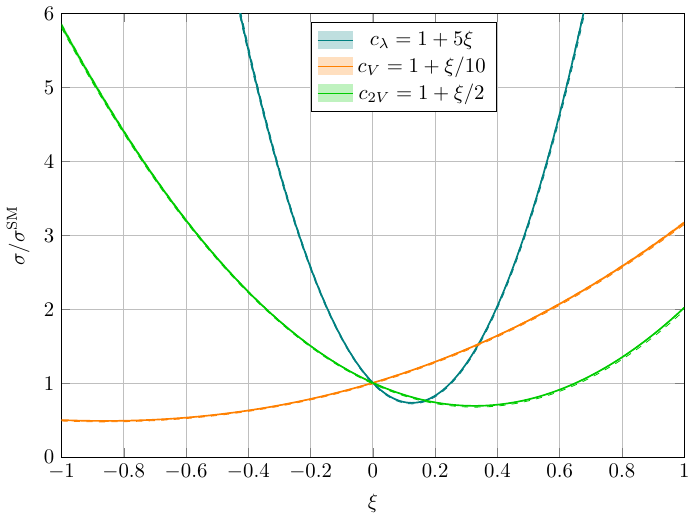}
    \end{subfigure}
    \begin{subfigure}[r]{0.49\textwidth}
        \includegraphics[width=\textwidth]{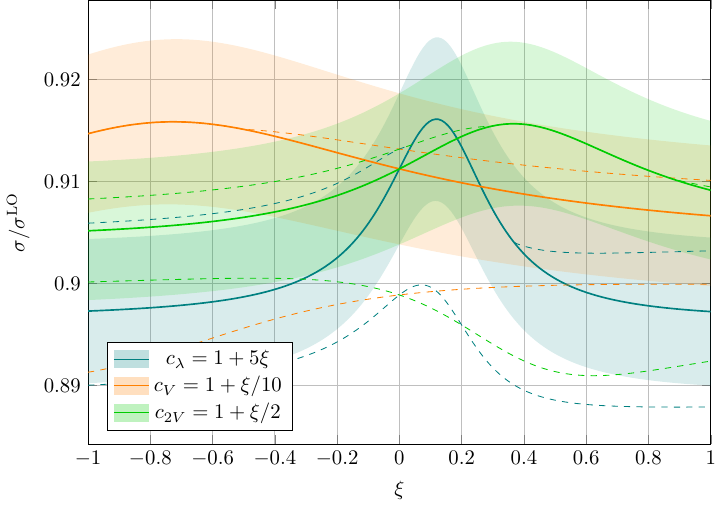}
    \end{subfigure}
    \caption{Ratio of the total cross section to the SM cross section at NLO (left) and to the LO cross section (right) for variations of a single coupling. The bands show the uncertainties resulting from the uncertainties of the fit coefficients, the dashed lines show the change due to the scale variation.\label{fig:spiderplot}}
\end{figure}

This parametrisation can now be used to investigate the dependence of the total cross sections on the anomalous couplings in the considered parameter space. Figure~\ref{fig:spiderplot} (left) shows the dependence of the cross section on a single anomalous coupling, where all other couplings are kept at the SM value. The impact of the different parameters varies strongly, with $c_\lambda$ having the largest impact, while $c_{V}$ has the smallest impact. However, this is mostly a result of the size of the respective parameter ranges. The vector boson couplings are much tighter constrained than the triple-Higgs coupling, allowing the latter to deviate much more from the SM value. Nonetheless, all couplings can significantly alter the cross section.
Maximising the function over the allowed parameter range leads to a value of about 27 times the SM cross section\footnote{To the best of our knowledge, the currently best experimental bound on the total cross section is 44 times the SM cross section~\cite{LHCHWG24}.}.
 
The right-hand side of figure~\ref{fig:spiderplot} shows the ratio of the total NLO cross section to the respective LO value, $\sigma_{NLO}/\sigma_{LO}$. 
We see that varying a single parameter over its whole parameter space impacts the ratio by less than 3\%. We can also observe the impact of the scale variation to be only relatively small, with the uncertainty due to the scale variation being of similar size as the uncertainties from the fit.

\begin{figure}
    \begin{subfigure}[l]{0.49\textwidth}
        \includegraphics[width=0.9\textwidth]{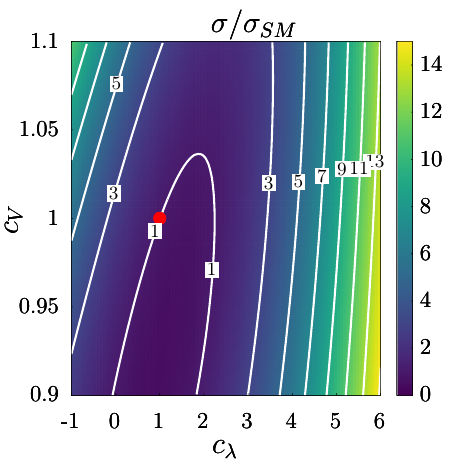}
    \end{subfigure}
    \begin{subfigure}[r]{0.49\textwidth}
        \includegraphics[width=0.9\textwidth]{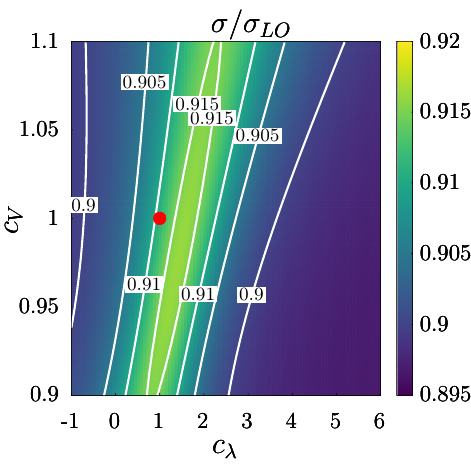}
    \end{subfigure}\\[-15pt]
    \begin{subfigure}[l]{0.49\textwidth}
        \includegraphics[width=0.9\textwidth]{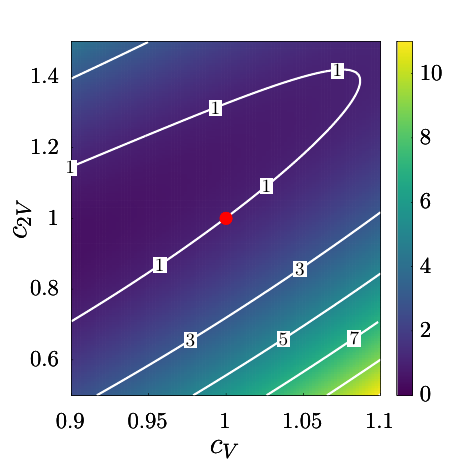}
    \end{subfigure}
    \begin{subfigure}[r]{0.49\textwidth}
        \includegraphics[width=0.9\textwidth]{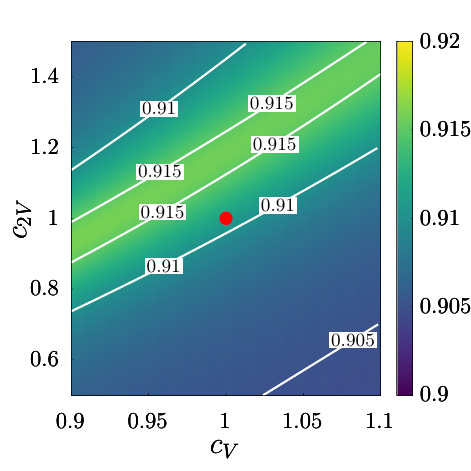}
    \end{subfigure}\\[-15pt]
    \begin{subfigure}[l]{0.49\textwidth}
        \includegraphics[width=0.9\textwidth]{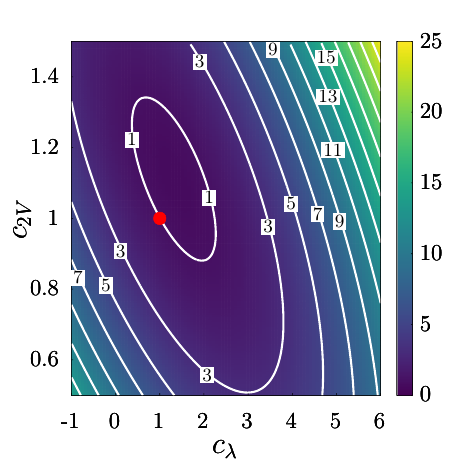}
    \end{subfigure}\hspace{0.15cm}
    \begin{subfigure}[r]{0.49\textwidth}
        \includegraphics[width=0.9\textwidth]{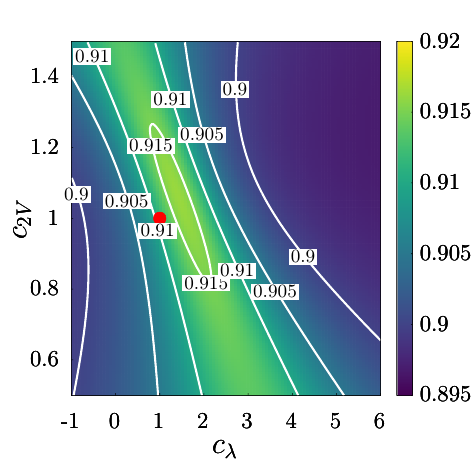}
    \end{subfigure}
    \caption{Ratio of the total cross section to the SM cross section at NLO (left column) and to the LO cross section (right column)
    in the $c_\lambda$-$c_V$ plane (first row), $c_V$-$c_{2V}$ plane (second row) and $c_\lambda$-$c_{2V}$ plane (third row). The red dot indicates the SM coupling values.}
    \label{fig:cv_clambda}
\end{figure}

So far, only variations of a single coupling were considered. In figure~\ref{fig:cv_clambda}, the simultaneous variation of two couplings is depicted. The left column contains the total NLO cross section relative to the SM, the right column again contains the ratio to the LO cross section, $\sigma_\mathrm{NLO}/\sigma_\mathrm{LO}$. For the slices shown in the coupling parameter space, the third (not shown) coupling is set to its SM value.
Variations in the ratio to LO still do not exceed the 3\% range, in agreement with the observations above.
However, the left column shows that there is a quite strong dependence of the total cross section on the values of the anomalous couplings. 
For large values of $c_\lambda$ and $c_{2V}$, the cross section can exceed the SM cross section by a factor of about 20, while for small values of $c_{V}$, the cross section can decrease.

\clearpage
\subsection{Differential Distributions}
\label{sec:differential}

Finally, we show differential results for four different observables:
the transverse momentum distribution $p_\perp^h$ of the Higgs bosons, the Higgs boson pair invariant mass $m_{hh}$, 
the pseudorapidity separation between the Higgs bosons, $\Delta\eta(h,h)$, and the $R$-separation between the Higgs bosons, 
$\Delta R(h,h)=\sqrt{\Delta\phi^2+\Delta\eta^2}$.
As the K-factor due to the NLO QCD corrections is mostly flat, we refrain from showing the SM LO and NLO curves and instead focus on the effects of the anomalous couplings on the NLO QCD results.
In particular, we show those benchmark points where a certain {\em combination} of anomalous couplings has a more pronounced effect on the shape of the distribution than varying any of these couplings individually.

\begin{figure}
    \begin{subfigure}[l]{0.49\textwidth}
        \includegraphics[width=\textwidth]{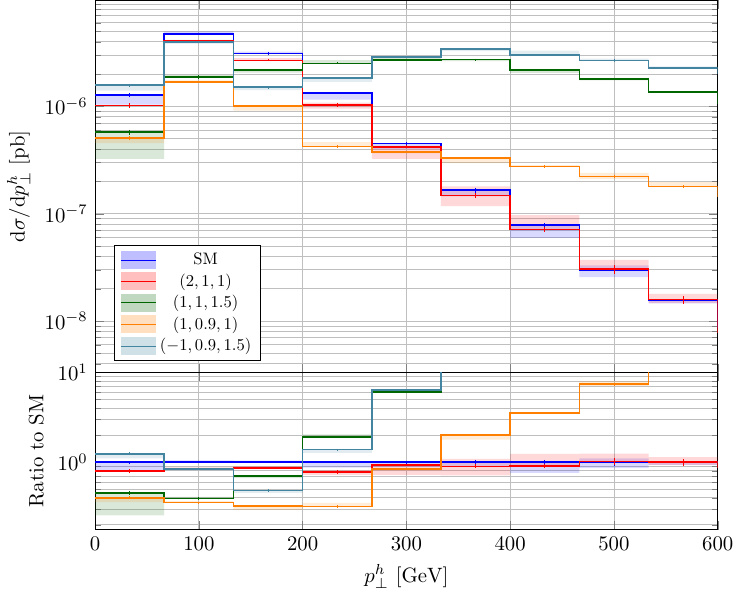}
    \end{subfigure}
    \begin{subfigure}[r]{0.49\textwidth}
        \includegraphics[width=\textwidth]{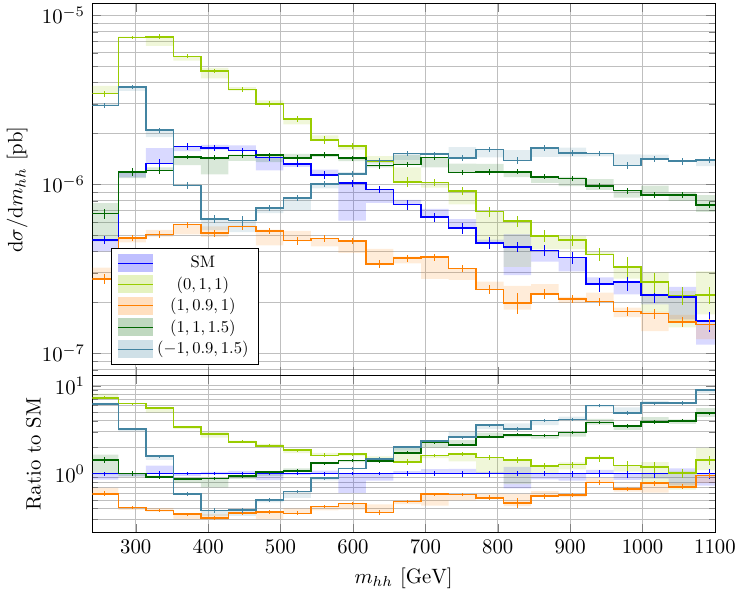}
    \end{subfigure}
    \caption{Transverse momentum distribution of (any of) the Higgs bosons (left) and invariant-mass distribution of the di-Higgs system (right) for selected benchmark points in $(c_\lambda,c_V,c_{2V})$-space of table \ref{tab:xsecs} at NLO.}
    \label{fig:differential_1}
  \end{figure}

In figure~\ref{fig:differential_1} we show the $p_\perp^h$ distribution for benchmark points 1, 4, 6 and 8 (left) and the $m_{hh}$ distribution for benchmark points 1, 3, 4, and 6 (right).
The  $p_\perp^h$ distribution clearly exhibits that variations of $c_\lambda$ alone do not have a dramatic effect, while changes in $c_V$ and $c_{2V}$ affect the delicate cancellations in the SM that guarantee unitarity and therefore have a large effect at high $p_\perp^h$. 
Unitarity constraints for the process $VV\to hh$ have been discussed in detail in ref.~\cite{Kilian:2018bhs}. Purely theoretical constraints can be useful to complement experimental constraints, 
see e.g. \cite{Cohen:2021ucp,Barducci:2023lqx} for recent studies.
This subject is beyond the scope of this paper, however.

The invariant mass of the di-Higgs system can show dramatic shape changes compared to the SM, both at low and high $m_{hh}$-values, where the changes at low $m_{hh}$ are induced by $c_\lambda$ being different from the SM value.
For benchmark point 1, destructive interference between different contributions leads to a characteristic peak-dip structure.
It is interesting to note that such a peak-dip structure is not present if $c_\lambda$ is only varied individually.

\begin{figure}
    \begin{subfigure}[l]{0.49\textwidth}
        \includegraphics[width=\textwidth]{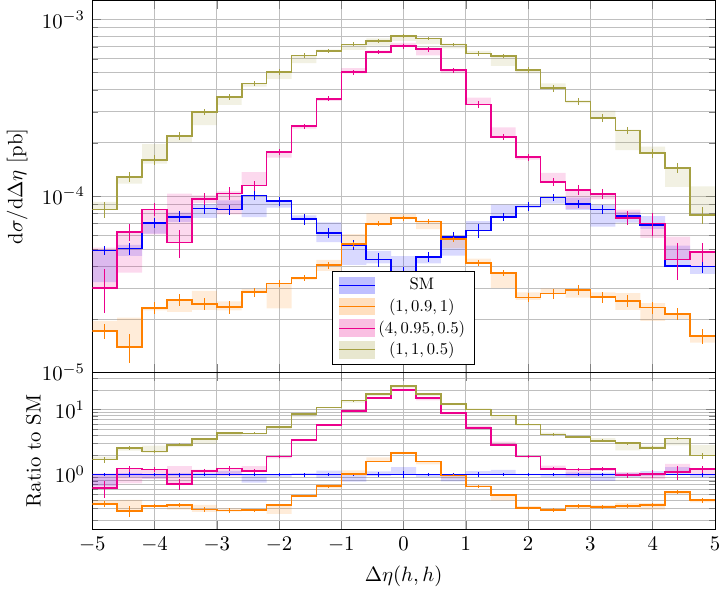}
    \end{subfigure}
    \begin{subfigure}[r]{0.49\textwidth}
        \includegraphics[width=\textwidth]{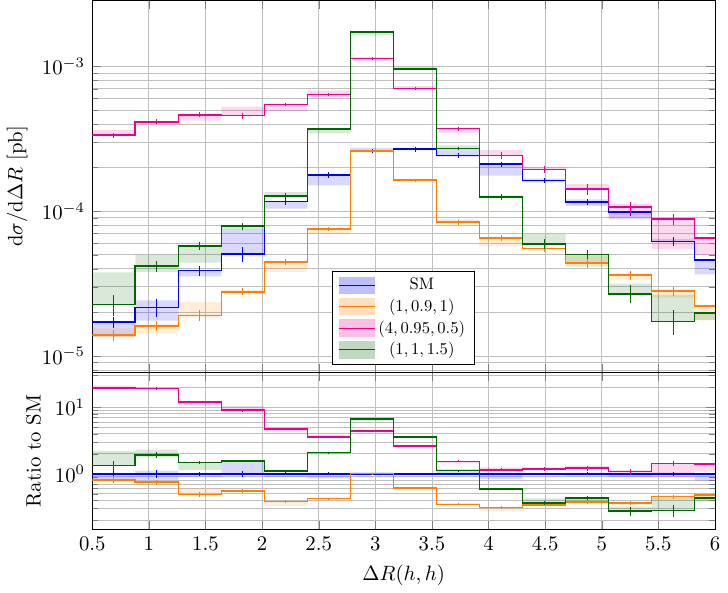}
    \end{subfigure}
    \caption{Pseudorapidity separation distribution (left) and $\Delta R$-separation distribution (right) of the Higgs pair for selected benchmark points in $(c_\lambda,c_V,c_{2V})$-space (see table~\ref{tab:xsecs}) at NLO.}
    \label{fig:differential_2}
\end{figure}

  We observed the largest shape changes in the $\Delta\eta(h,h)$ distribution, shown in figure~\ref{fig:differential_2} (left), as well as some interesting features in the $\Delta R(h,h)$ distribution (figure~\ref{fig:differential_2} right).
  The former distribution is shown for the benchmark points 4, 5 and 10, the latter for 4, 6 and 10.
  In the SM, most Higgs boson pairs are produced with a pseudorapidity separation of about 2.5 and a clear local minimum at $\Delta\eta(h,h)=0$.
  However, this dip becomes a local maximum for the shown benchmark points, for example forming a distinct peak for benchmark point 10, which is $(c_\lambda,c_V,c_{2V})=(4,0.95,0.5)$.  
  This behaviour can be understood by considering the diagrams in figure~\ref{fig:VBF_LO_sub}. The contribution of the leftmost diagram is enhanced for large self-couplings, leading to an excess in less well separated Higgs bosons, since they originate from the same triple Higgs three-point vertex.
  
The changes in the pseudo-rapidity separation also influence the $R$-separation $\Delta R$  of the Higgs bosons. This distribution is also skewed towards smaller $\Delta R$ values for large values of $c_\lambda$, due to the dominance of the contribution from diagrams with an s-channel Higgs propagator. However, due to a simultaneous increase in $\Delta\phi$ for anomalous values of the Higgs-vector boson couplings, another interesting feature can be observed: increasing the Higgs boson self-coupling only increases the abundance of low $\Delta R$ events, while for variations of the other two couplings, a peak develops around $\Delta R\sim \pi$. The large--$\Delta R$ region is much less affected by changes in the couplings.

\clearpage
\section{Conclusions}
\label{sec:conclusions}

We have calculated the NLO QCD corrections to Higgs boson pair production in vector boson fusion and included the leading anomalous couplings within non-linear Effective Field Theory (HEFT).
We explained in detail the derivation of the HEFT Lagrangian for this process and the contributing diagrams. We did not use the structure function approximation; diagrams of $u$- and $s$-channel type (``Higgs-Strahlung topologies") are also included. 

Our results are based on an automated \whizard-\gosam{} interface that can be used for user-defined NLO calculations for both hadronic or leptonic collisions, including the option to use UFO model files and flexible setups for calculations beyond the SM.
The new releases of both tools are close to completion.
This setup also allows for parton showering, however we did not add a shower in the present work.

As is well known, the NLO QCD corrections in the SM reduce the cross section by about 7\% with a relatively flat K-factor. 
The K-factors relative to the SM K-factors also do not vary by more then about 3\% when introducing anomalous couplings, which can be understood from the fact that the QCD corrections of ``structure function type" are dominating, while the anomalous couplings affect the electroweak part of the amplitude.
We have parametrised the total cross section as a polynomial in the anomalous couplings and fitted the coefficients based on our NLO calculation. This offers a flexible representation of the cross section that allows fast evaluation for arbitrary values of the couplings. Based on this parametrisation, we have shown heat maps for slices of the $(c_\lambda,c_V,c_{2V})$-parameter space illustrating the ratio to the SM cross section and the respective LO cross section.

In the anomalous coupling parameter space, we defined 11 benchmark points, oriented at the current experimental constraints, which lead to characteristic shapes in the $p_\perp^h$, $m_{hh}$, $\Delta\eta(h,h)$ or $\Delta R(h,h)$ distributions.
It is not surprising that modifications of $c_V$ or $c_{2V}$ can have large effects in the high-energy tails of the distributions, as the SM-type unitarity cancellations are spoiled and may only set in later through a particular UV completion.
A feature that is noteworthy is the fact that combinations of anomalous couplings can lead to characteristic shapes that would not occur if any of the couplings was varied in isolation. This feature also can be observed in phase space regions that are experimentally more accessible than the tails of distributions, such as the central region of the pseudo-rapidity difference $\Delta\eta(h,h)$.
Such characteristic shapes could signpost physics beyond the SM even with moderate experimental statistics.

\subsection*{Note added}

During completion of this paper, we became aware of similar work~\cite{Jager:2025isz}, where an NLO+PS calculation of this process is presented in the {\tt Powheg-Box-V2} framework, also accounting for anomalous couplings, together with an upgrade of the program {\tt proVBFHH}~\cite{proVBFHH}. However, the two calculations are in large parts complementary: The work of \cite{Jager:2025isz} focuses on the comparison between the NNLO results and the NLO+PS results and uses the structure function approximation. Our calculation does not use this approximation and focuses on the derivation of the leading operators in HEFT and how the corresponding anomalous couplings affect the shape of distributions if varied simultaneously.
We leave a more detailed comparison for future work. 

\section*{Acknowledgements}

We would like to thank Alexander Karlberg, Barbara J\"ager and Simon Reinhardt for sharing their draft with us before publication. We also would like to thank Jannis Lang for work on the EFT aspects of \gosam{} and J\"urgen Reuter and Wolfgang Kilian for input concerning the \whizard-\gosam{} interface.
This research was supported by the Deutsche Forschungsgemeinschaft (DFG, German Research Foundation) under grant 396021762 - TRR 257. The authors also acknowledge support by the state of Baden-Württemberg through bwHPC.


\bibliographystyle{JHEP}
\bibliography{vbf_heft}

\providecommand{\href}[2]{#2}\begingroup\raggedright\begin{thebibliography}{10}

\bibitem{ATLAS:2024ish}
{\scshape ATLAS} collaboration, G.~Aad et~al., \emph{{Combination of Searches
  for Higgs Boson Pair Production in pp Collisions at $\sqrt{s}=13$\,TeV with
  the ATLAS Detector}},
  \href{https://doi.org/10.1103/PhysRevLett.133.101801}{\emph{Phys. Rev. Lett.}
  {\bfseries 133} (2024) 101801}
  [\href{https://arxiv.org/abs/2406.09971}{{\ttfamily 2406.09971}}].

\bibitem{CMS:2024awa}
{\scshape CMS} collaboration, A.~Hayrapetyan et~al., \emph{{Constraints on the
  Higgs boson self-coupling from the combination of single and double Higgs
  boson production in proton-proton collisions at $\sqrt{s}=13$\,TeV}},
  \href{https://doi.org/10.1016/j.physletb.2024.139210}{\emph{Phys. Lett. B}
  {\bfseries 861} (2025) 139210}
  [\href{https://arxiv.org/abs/2407.13554}{{\ttfamily 2407.13554}}].

\bibitem{ATLAS:2024lsk}
{\scshape ATLAS} collaboration, G.~Aad et~al., \emph{{Search for pair
  production of boosted Higgs bosons via vector-boson fusion in the
  $b\bar{b}b\bar{b}$ final state using pp collisions at $\sqrt{s}=13$\,TeV with
  the ATLAS detector}},
  \href{https://doi.org/10.1016/j.physletb.2024.139007}{\emph{Phys. Lett. B}
  {\bfseries 858} (2024) 139007}
  [\href{https://arxiv.org/abs/2404.17193}{{\ttfamily 2404.17193}}].

\bibitem{CMS-PAS-HIG-20-011}
{\scshape CMS} collaboration, \emph{{Combination of searches for nonresonant
  Higgs boson pair production in proton-proton collisions at sqrt(s) = 13
  TeV}},  tech. rep., CERN, Geneva, 2024.

\bibitem{Figy:2008zd}
T.~Figy, \emph{{Next-to-leading order QCD corrections to light Higgs Pair
  production via vector boson fusion}},
  \href{https://doi.org/10.1142/S0217732308028181}{\emph{Mod. Phys. Lett. A}
  {\bfseries 23} (2008) 1961}
  [\href{https://arxiv.org/abs/0806.2200}{{\ttfamily 0806.2200}}].

\bibitem{Baglio:2012np}
J.~Baglio, A.~Djouadi, R.~Gr\"ober, M.~M. M\"uhlleitner, J.~Quevillon and
  M.~Spira, \emph{{The measurement of the Higgs self-coupling at the LHC:
  theoretical status}},
  \href{https://doi.org/10.1007/JHEP04(2013)151}{\emph{JHEP} {\bfseries 04}
  (2013) 151} [\href{https://arxiv.org/abs/1212.5581}{{\ttfamily 1212.5581}}].

\bibitem{Frederix:2014hta}
R.~Frederix, S.~Frixione, V.~Hirschi, F.~Maltoni, O.~Mattelaer, P.~Torrielli
  et~al., \emph{{Higgs pair production at the LHC with NLO and parton-shower
  effects}}, \href{https://doi.org/10.1016/j.physletb.2014.03.026}{\emph{Phys.
  Lett. B} {\bfseries 732} (2014) 142}
  [\href{https://arxiv.org/abs/1401.7340}{{\ttfamily 1401.7340}}].

\bibitem{Arnold:2008rz}
K.~Arnold et~al., \emph{{VBFNLO: A Parton level Monte Carlo for processes with
  electroweak bosons}},
  \href{https://doi.org/10.1016/j.cpc.2009.03.006}{\emph{Comput. Phys. Commun.}
  {\bfseries 180} (2009) 1661}
  [\href{https://arxiv.org/abs/0811.4559}{{\ttfamily 0811.4559}}].

\bibitem{Baglio:2024gyp}
J.~Baglio et~al., \emph{{Release note: VBFNLO 3.0}},
  \href{https://doi.org/10.1140/epjc/s10052-024-13336-x}{\emph{Eur. Phys. J. C}
  {\bfseries 84} (2024) 1003}
  [\href{https://arxiv.org/abs/2405.06990}{{\ttfamily 2405.06990}}].

\bibitem{Alwall:2014hca}
J.~Alwall, R.~Frederix, S.~Frixione, V.~Hirschi, F.~Maltoni, O.~Mattelaer
  et~al., \emph{{The automated computation of tree-level and next-to-leading
  order differential cross sections, and their matching to parton shower
  simulations}}, \href{https://doi.org/10.1007/JHEP07(2014)079}{\emph{JHEP}
  {\bfseries 07} (2014) 079} [\href{https://arxiv.org/abs/1405.0301}{{\ttfamily
  1405.0301}}].

\bibitem{Ling:2014sne}
L.-S. Ling, R.-Y. Zhang, W.-G. Ma, L.~Guo, W.-H. Li and X.-Z. Li, \emph{{NNLO
  QCD corrections to Higgs pair production via vector boson fusion at hadron
  colliders}}, \href{https://doi.org/10.1103/PhysRevD.89.073001}{\emph{Phys.
  Rev. D} {\bfseries 89} (2014) 073001}
  [\href{https://arxiv.org/abs/1401.7754}{{\ttfamily 1401.7754}}].

\bibitem{Dreyer:2018rfu}
F.~A. Dreyer and A.~Karlberg, \emph{{Fully differential Vector-Boson Fusion
  Higgs Pair Production at Next-to-Next-to-Leading Order}},
  \href{https://doi.org/10.1103/PhysRevD.99.074028}{\emph{Phys. Rev. D}
  {\bfseries 99} (2019) 074028}
  [\href{https://arxiv.org/abs/1811.07918}{{\ttfamily 1811.07918}}].

\bibitem{Dreyer:2018qbw}
F.~A. Dreyer and A.~Karlberg, \emph{{Vector-Boson Fusion Higgs Pair Production
  at N$^3$LO}}, \href{https://doi.org/10.1103/PhysRevD.98.114016}{\emph{Phys.
  Rev. D} {\bfseries 98} (2018) 114016}
  [\href{https://arxiv.org/abs/1811.07906}{{\ttfamily 1811.07906}}].

\bibitem{Dreyer:2020xaj}
F.~A. Dreyer, A.~Karlberg, J.-N. Lang and M.~Pellen, \emph{{Precise predictions
  for double-Higgs production via vector-boson fusion}},
  \href{https://doi.org/10.1140/epjc/s10052-020-08610-7}{\emph{Eur. Phys. J. C}
  {\bfseries 80} (2020) 1037}
  [\href{https://arxiv.org/abs/2005.13341}{{\ttfamily 2005.13341}}].

\bibitem{Dreyer:2020urf}
F.~A. Dreyer, A.~Karlberg and L.~Tancredi, \emph{{On the impact of
  non-factorisable corrections in VBF single and double Higgs production}},
  \href{https://doi.org/10.1007/JHEP10(2020)131}{\emph{JHEP} {\bfseries 10}
  (2020) 131} [\href{https://arxiv.org/abs/2005.11334}{{\ttfamily
  2005.11334}}].

\bibitem{proVBFHH}
A.~Karlberg, ``{proVBFHH}.''
  \url{https://github.com/alexanderkarlberg/proVBFH}.

\bibitem{Dolan:2013rja}
M.~J. Dolan, C.~Englert, N.~Greiner and M.~Spannowsky, \emph{{Further on up the
  road: $hhjj$ production at the LHC}},
  \href{https://doi.org/10.1103/PhysRevLett.112.101802}{\emph{Phys. Rev. Lett.}
  {\bfseries 112} (2014) 101802}
  [\href{https://arxiv.org/abs/1310.1084}{{\ttfamily 1310.1084}}].

\bibitem{Dolan:2015zja}
M.~J. Dolan, C.~Englert, N.~Greiner, K.~Nordstrom and M.~Spannowsky,
  \emph{{$hhjj$ production at the LHC}},
  \href{https://doi.org/10.1140/epjc/s10052-015-3622-3}{\emph{Eur. Phys. J. C}
  {\bfseries 75} (2015) 387}
  [\href{https://arxiv.org/abs/1506.08008}{{\ttfamily 1506.08008}}].

\bibitem{Bishara:2016kjn}
F.~Bishara, R.~Contino and J.~Rojo, \emph{{Higgs pair production in
  vector-boson fusion at the LHC and beyond}},
  \href{https://doi.org/10.1140/epjc/s10052-017-5037-9}{\emph{Eur. Phys. J. C}
  {\bfseries 77} (2017) 481}
  [\href{https://arxiv.org/abs/1611.03860}{{\ttfamily 1611.03860}}].

\bibitem{Ling:2017teo}
L.-S. Ling, R.-Y. Zhang, W.-G. Ma, X.-Z. Li, L.~Guo and S.-M. Wang,
  \emph{{Dimension-six operators in Higgs boson pair production via
  vector-boson fusion at the LHC}},
  \href{https://doi.org/10.1103/PhysRevD.96.055006}{\emph{Phys. Rev. D}
  {\bfseries 96} (2017) 055006}
  [\href{https://arxiv.org/abs/1708.04785}{{\ttfamily 1708.04785}}].

\bibitem{Arganda:2018ftn}
E.~Arganda, C.~Garcia-Garcia and M.~J. Herrero, \emph{{Probing the Higgs
  self-coupling through double Higgs production in vector boson scattering at
  the LHC}}, \href{https://doi.org/10.1016/j.nuclphysb.2019.114687}{\emph{Nucl.
  Phys. B} {\bfseries 945} (2019) 114687}
  [\href{https://arxiv.org/abs/1807.09736}{{\ttfamily 1807.09736}}].

\bibitem{Araz:2020zyh}
J.~Y. Araz, S.~Banerjee, R.~S. Gupta and M.~Spannowsky, \emph{{Precision SMEFT
  bounds from the VBF Higgs at high transverse momentum}},
  \href{https://doi.org/10.1007/JHEP04(2021)125}{\emph{JHEP} {\bfseries 04}
  (2021) 125} [\href{https://arxiv.org/abs/2011.03555}{{\ttfamily
  2011.03555}}].

\bibitem{Kilian:2018bhs}
W.~Kilian, S.~Sun, Q.-S. Yan, X.~Zhao and Z.~Zhao, \emph{{Multi-Higgs boson
  production and unitarity in vector-boson fusion at future hadron colliders}},
  \href{https://doi.org/10.1103/PhysRevD.101.076012}{\emph{Phys. Rev. D}
  {\bfseries 101} (2020) 076012}
  [\href{https://arxiv.org/abs/1808.05534}{{\ttfamily 1808.05534}}].

\bibitem{Kilian:2021whd}
W.~Kilian, S.~Sun, Q.-S. Yan, X.~Zhao and Z.~Zhao, \emph{{Highly Boosted Higgs
  Bosons and Unitarity in Vector-Boson Fusion at Future Hadron Colliders}},
  \href{https://doi.org/10.1007/JHEP05(2021)198}{\emph{JHEP} {\bfseries 05}
  (2021) 198} [\href{https://arxiv.org/abs/2101.12537}{{\ttfamily
  2101.12537}}].

\bibitem{Gomez-Ambrosio:2022qsi}
R.~G\'omez-Ambrosio, F.~J. Llanes-Estrada, A.~Salas-Bern\'ardez and J.~J.
  Sanz-Cillero, \emph{{Distinguishing electroweak EFTs with $W_LW_L\to n\,h$}},
  \href{https://doi.org/10.1103/PhysRevD.106.053004}{\emph{Phys. Rev. D}
  {\bfseries 106} (2022) 053004}
  [\href{https://arxiv.org/abs/2204.01763}{{\ttfamily 2204.01763}}].

\bibitem{Gomez-Ambrosio:2022why}
R.~G\'omez-Ambrosio, F.~J. Llanes-Estrada, A.~Salas-Bern\'ardez and J.~J.
  Sanz-Cillero, \emph{{SMEFT is falsifiable through multi-Higgs measurements
  (even in the absence of new light particles)}},
  \href{https://doi.org/10.1088/1572-9494/ace95e}{\emph{Commun. Theor. Phys.}
  {\bfseries 75} (2023) 095202}
  [\href{https://arxiv.org/abs/2207.09848}{{\ttfamily 2207.09848}}].

\bibitem{Herrero:2022krh}
M.~J. Herrero and R.~A. Morales, \emph{{One-loop corrections for WW to HH in
  Higgs EFT with the electroweak chiral Lagrangian}},
  \href{https://doi.org/10.1103/PhysRevD.106.073008}{\emph{Phys. Rev. D}
  {\bfseries 106} (2022) 073008}
  [\href{https://arxiv.org/abs/2208.05900}{{\ttfamily 2208.05900}}].

\bibitem{Anisha:2024ljc}
Anisha, D.~Domenech, C.~Englert, M.~J. Herrero and R.~A. Morales,
  \emph{{Bosonic multi-Higgs correlations beyond leading order}},
  \href{https://doi.org/10.1103/PhysRevD.110.095016}{\emph{Phys. Rev. D}
  {\bfseries 110} (2024) 095016}
  [\href{https://arxiv.org/abs/2405.05385}{{\ttfamily 2405.05385}}].

\bibitem{Anisha:2024ryj}
Anisha, D.~Domenech, C.~Englert, M.~J. Herrero and R.~A. Morales, \emph{{HEFT's
  appraisal of triple (versus double) Higgs weak boson fusion}},
  \href{https://arxiv.org/abs/2407.20706}{{\ttfamily 2407.20706}}.

\bibitem{Feruglio:1992wf}
F.~Feruglio, \emph{{The Chiral approach to the electroweak interactions}},
  \href{https://doi.org/10.1142/S0217751X93001946}{\emph{Int. J. Mod. Phys. A}
  {\bfseries 8} (1993) 4937}
  [\href{https://arxiv.org/abs/hep-ph/9301281}{{\ttfamily hep-ph/9301281}}].

\bibitem{Bagger:1993zf}
J.~Bagger, V.~D. Barger, K.-m. Cheung, J.~F. Gunion, T.~Han, G.~A. Ladinsky
  et~al., \emph{{The Strongly interacting W W system: Gold plated modes}},
  \href{https://doi.org/10.1103/PhysRevD.49.1246}{\emph{Phys. Rev. D}
  {\bfseries 49} (1994) 1246}
  [\href{https://arxiv.org/abs/hep-ph/9306256}{{\ttfamily hep-ph/9306256}}].

\bibitem{Koulovassilopoulos:1993pw}
V.~Koulovassilopoulos and R.~S. Chivukula, \emph{{The Phenomenology of a
  nonstandard Higgs boson in W(L) W(L) scattering}},
  \href{https://doi.org/10.1103/PhysRevD.50.3218}{\emph{Phys. Rev. D}
  {\bfseries 50} (1994) 3218}
  [\href{https://arxiv.org/abs/hep-ph/9312317}{{\ttfamily hep-ph/9312317}}].

\bibitem{Burgess:1999ha}
C.~P. Burgess, J.~Matias and M.~Pospelov, \emph{{A Higgs or not a Higgs? What
  to do if you discover a new scalar particle}},
  \href{https://doi.org/10.1142/S0217751X02009813}{\emph{Int. J. Mod. Phys. A}
  {\bfseries 17} (2002) 1841}
  [\href{https://arxiv.org/abs/hep-ph/9912459}{{\ttfamily hep-ph/9912459}}].

\bibitem{Wang:2006im}
L.-M. Wang and Q.~Wang, \emph{{Nonstandard Higgs in electroweak chiral
  Lagrangian}},  \href{https://arxiv.org/abs/hep-ph/0605104}{{\ttfamily
  hep-ph/0605104}}.

\bibitem{Grinstein:2007iv}
B.~Grinstein and M.~Trott, \emph{{A Higgs-Higgs bound state due to new physics
  at a TeV}}, \href{https://doi.org/10.1103/PhysRevD.76.073002}{\emph{Phys.
  Rev. D} {\bfseries 76} (2007) 073002}
  [\href{https://arxiv.org/abs/0704.1505}{{\ttfamily 0704.1505}}].

\bibitem{Alonso:2012px}
R.~Alonso, M.~B. Gavela, L.~Merlo, S.~Rigolin and J.~Yepes, \emph{{The
  Effective Chiral Lagrangian for a Light Dynamical ''Higgs Particle''}},
  \href{https://doi.org/10.1016/j.physletb.2013.04.037}{\emph{Phys. Lett. B}
  {\bfseries 722} (2013) 330}
  [\href{https://arxiv.org/abs/1212.3305}{{\ttfamily 1212.3305}}].

\bibitem{Buchalla:2012qq}
G.~Buchalla and O.~Cata, \emph{{Effective Theory of a Dynamically Broken
  Electroweak Standard Model at NLO}},
  \href{https://doi.org/10.1007/JHEP07(2012)101}{\emph{JHEP} {\bfseries 07}
  (2012) 101} [\href{https://arxiv.org/abs/1203.6510}{{\ttfamily 1203.6510}}].

\bibitem{Buchalla:2013rka}
G.~Buchalla, O.~Cat\`a and C.~Krause, \emph{{Complete Electroweak Chiral
  Lagrangian with a Light Higgs at NLO}},
  \href{https://doi.org/10.1016/j.nuclphysb.2014.01.018}{\emph{Nucl. Phys. B}
  {\bfseries 880} (2014) 552}
  [\href{https://arxiv.org/abs/1307.5017}{{\ttfamily 1307.5017}}].

\bibitem{Degrande:2020evl}
C.~Degrande, G.~Durieux, F.~Maltoni, K.~Mimasu, E.~Vryonidou and C.~Zhang,
  \emph{{Automated one-loop computations in the standard model effective field
  theory}}, \href{https://doi.org/10.1103/PhysRevD.103.096024}{\emph{Phys. Rev.
  D} {\bfseries 103} (2021) 096024}
  [\href{https://arxiv.org/abs/2008.11743}{{\ttfamily 2008.11743}}].

\bibitem{Cullen:2011ac}
{\scshape GoSam} collaboration, G.~Cullen, N.~Greiner, G.~Heinrich, G.~Luisoni,
  P.~Mastrolia, G.~Ossola et~al., \emph{{Automated One-Loop Calculations with
  GoSam}}, \href{https://doi.org/10.1140/epjc/s10052-012-1889-1}{\emph{Eur.
  Phys. J. C} {\bfseries 72} (2012) 1889}
  [\href{https://arxiv.org/abs/1111.2034}{{\ttfamily 1111.2034}}].

\bibitem{GoSam:2014iqq}
{\scshape GoSam} collaboration, G.~Cullen et~al., \emph{{GoSam-2.0: a tool for
  automated one-loop calculations within the Standard Model and beyond}},
  \href{https://doi.org/10.1140/epjc/s10052-014-3001-5}{\emph{Eur. Phys. J. C}
  {\bfseries 74} (2014) 3001}
  [\href{https://arxiv.org/abs/1404.7096}{{\ttfamily 1404.7096}}].

\bibitem{toappear}
J.~Braun et~al., \emph{{GoSam-3.0, to appear}}, .

\bibitem{Moretti:2001zz}
M.~Moretti, T.~Ohl and J.~Reuter, \emph{{O'Mega: An Optimizing matrix element
  generator}},  \href{https://arxiv.org/abs/hep-ph/0102195}{{\ttfamily
  hep-ph/0102195}}.

\bibitem{Kilian:2007gr}
W.~Kilian, T.~Ohl and J.~Reuter, \emph{{WHIZARD: Simulating Multi-Particle
  Processes at LHC and ILC}},
  \href{https://doi.org/10.1140/epjc/s10052-011-1742-y}{\emph{Eur. Phys. J. C}
  {\bfseries 71} (2011) 1742}
  [\href{https://arxiv.org/abs/0708.4233}{{\ttfamily 0708.4233}}].

\bibitem{Binoth:2010xt}
T.~Binoth et~al., \emph{{A Proposal for a Standard Interface between Monte
  Carlo Tools and One-Loop Programs}},
  \href{https://doi.org/10.1016/j.cpc.2010.05.016}{\emph{Comput. Phys. Commun.}
  {\bfseries 181} (2010) 1612}
  [\href{https://arxiv.org/abs/1001.1307}{{\ttfamily 1001.1307}}].

\bibitem{Alioli:2013nda}
S.~Alioli et~al., \emph{{Update of the Binoth Les Houches Accord for a standard
  interface between Monte Carlo tools and one-loop programs}},
  \href{https://doi.org/10.1016/j.cpc.2013.10.020}{\emph{Comput. Phys. Commun.}
  {\bfseries 185} (2014) 560}
  [\href{https://arxiv.org/abs/1308.3462}{{\ttfamily 1308.3462}}].

\bibitem{Degrande:2011ua}
C.~Degrande, C.~Duhr, B.~Fuks, D.~Grellscheid, O.~Mattelaer and T.~Reiter,
  \emph{{UFO - The Universal FeynRules Output}},
  \href{https://doi.org/10.1016/j.cpc.2012.01.022}{\emph{Comput. Phys. Commun.}
  {\bfseries 183} (2012) 1201}
  [\href{https://arxiv.org/abs/1108.2040}{{\ttfamily 1108.2040}}].

\bibitem{Darme:2023jdn}
L.~Darm\'e et~al., \emph{{UFO 2.0: the ``Universal Feynman Output" format}},
  \href{https://doi.org/10.1140/epjc/s10052-023-11780-9}{\emph{Eur. Phys. J. C}
  {\bfseries 83} (2023) 631}
  [\href{https://arxiv.org/abs/2304.09883}{{\ttfamily 2304.09883}}].

\bibitem{Buchmuller:1985jz}
W.~Buchm{\"u}ller and D.~Wyler, \emph{{Effective Lagrangian Analysis of New
  Interactions and Flavor Conservation}},
  \href{https://doi.org/10.1016/0550-3213(86)90262-2}{\emph{Nucl. Phys. B}
  {\bfseries 268} (1986) 621}.

\bibitem{Grzadkowski:2010es}
B.~Grzadkowski, M.~Iskrzynski, M.~Misiak and J.~Rosiek, \emph{{Dimension-Six
  Terms in the Standard Model Lagrangian}},
  \href{https://doi.org/10.1007/JHEP10(2010)085}{\emph{JHEP} {\bfseries 10}
  (2010) 085} [\href{https://arxiv.org/abs/1008.4884}{{\ttfamily 1008.4884}}].

\bibitem{Brivio:2017vri}
I.~Brivio and M.~Trott, \emph{{The Standard Model as an Effective Field
  Theory}}, \href{https://doi.org/10.1016/j.physrep.2018.11.002}{\emph{Phys.
  Rept.} {\bfseries 793} (2019) 1}
  [\href{https://arxiv.org/abs/1706.08945}{{\ttfamily 1706.08945}}].

\bibitem{Isidori:2023pyp}
G.~Isidori, F.~Wilsch and D.~Wyler, \emph{{The standard model effective field
  theory at work}},
  \href{https://doi.org/10.1103/RevModPhys.96.015006}{\emph{Rev. Mod. Phys.}
  {\bfseries 96} (2024) 015006}
  [\href{https://arxiv.org/abs/2303.16922}{{\ttfamily 2303.16922}}].

\bibitem{Arzt:1994gp}
C.~Arzt, M.~B. Einhorn and J.~Wudka, \emph{{Patterns of deviation from the
  standard model}},
  \href{https://doi.org/10.1016/0550-3213(94)00336-D}{\emph{Nucl. Phys. B}
  {\bfseries 433} (1995) 41}
  [\href{https://arxiv.org/abs/hep-ph/9405214}{{\ttfamily hep-ph/9405214}}].

\bibitem{Craig:2019wmo}
N.~Craig, M.~Jiang, Y.-Y. Li and D.~Sutherland, \emph{{Loops and Trees in
  Generic EFTs}}, \href{https://doi.org/10.1007/JHEP08(2020)086}{\emph{JHEP}
  {\bfseries 08} (2020) 086}
  [\href{https://arxiv.org/abs/2001.00017}{{\ttfamily 2001.00017}}].

\bibitem{Buchalla:2022vjp}
G.~Buchalla, G.~Heinrich, C.~M\"uller-Salditt and F.~Pandler, \emph{{Loop
  counting matters in SMEFT}},
  \href{https://doi.org/10.21468/SciPostPhys.15.3.088}{\emph{SciPost Phys.}
  {\bfseries 15} (2023) 088}
  [\href{https://arxiv.org/abs/2204.11808}{{\ttfamily 2204.11808}}].

\bibitem{Buchalla:2013eza}
G.~Buchalla, O.~Cat\'a and C.~Krause, \emph{{On the Power Counting in Effective
  Field Theories}},
  \href{https://doi.org/10.1016/j.physletb.2014.02.015}{\emph{Phys. Lett. B}
  {\bfseries 731} (2014) 80} [\href{https://arxiv.org/abs/1312.5624}{{\ttfamily
  1312.5624}}].

\bibitem{Sun:2022ssa}
H.~Sun, M.-L. Xiao and J.-H. Yu, \emph{{Complete NLO operators in the Higgs
  effective field theory}},
  \href{https://doi.org/10.1007/JHEP05(2023)043}{\emph{JHEP} {\bfseries 05}
  (2023) 043} [\href{https://arxiv.org/abs/2206.07722}{{\ttfamily
  2206.07722}}].

\bibitem{Graf:2022rco}
L.~Gr\'af, B.~Henning, X.~Lu, T.~Melia and H.~Murayama, \emph{{Hilbert series,
  the Higgs mechanism, and HEFT}},
  \href{https://doi.org/10.1007/JHEP02(2023)064}{\emph{JHEP} {\bfseries 02}
  (2023) 064} [\href{https://arxiv.org/abs/2211.06275}{{\ttfamily
  2211.06275}}].

\bibitem{Alonso:2017tdy}
R.~Alonso, K.~Kanshin and S.~Saa, \emph{{Renormalization group evolution of
  Higgs effective field theory}},
  \href{https://doi.org/10.1103/PhysRevD.97.035010}{\emph{Phys. Rev. D}
  {\bfseries 97} (2018) 035010}
  [\href{https://arxiv.org/abs/1710.06848}{{\ttfamily 1710.06848}}].

\bibitem{Buchalla:2017jlu}
G.~Buchalla, O.~Cata, A.~Celis, M.~Knecht and C.~Krause, \emph{{Complete
  One-Loop Renormalization of the Higgs-Electroweak Chiral Lagrangian}},
  \href{https://doi.org/10.1016/j.nuclphysb.2018.01.009}{\emph{Nucl. Phys. B}
  {\bfseries 928} (2018) 93}
  [\href{https://arxiv.org/abs/1710.06412}{{\ttfamily 1710.06412}}].

\bibitem{Buchalla:2020kdh}
G.~Buchalla, O.~Cat\`a, A.~Celis, M.~Knecht and C.~Krause,
  \emph{{Higgs-electroweak chiral Lagrangian: One-loop renormalization group
  equations}}, \href{https://doi.org/10.1103/PhysRevD.104.076005}{\emph{Phys.
  Rev. D} {\bfseries 104} (2021) 076005}
  [\href{https://arxiv.org/abs/2004.11348}{{\ttfamily 2004.11348}}].

\bibitem{Peskin:1991sw}
M.~E. Peskin and T.~Takeuchi, \emph{{Estimation of oblique electroweak
  corrections}}, \href{https://doi.org/10.1103/PhysRevD.46.381}{\emph{Phys.
  Rev. D} {\bfseries 46} (1992) 381}.

\bibitem{deBlas:2021wap}
J.~de~Blas, M.~Ciuchini, E.~Franco, A.~Goncalves, S.~Mishima, M.~Pierini
  et~al., \emph{{Global analysis of electroweak data in the Standard Model}},
  \href{https://doi.org/10.1103/PhysRevD.106.033003}{\emph{Phys. Rev. D}
  {\bfseries 106} (2022) 033003}
  [\href{https://arxiv.org/abs/2112.07274}{{\ttfamily 2112.07274}}].

\bibitem{LHCHiggsCrossSectionWorkingGroup:2012nn}
{\scshape LHC Higgs Cross Section Working Group} collaboration, A.~David,
  A.~Denner, M.~Duehrssen, M.~Grazzini, C.~Grojean, G.~Passarino et~al.,
  \emph{{LHC HXSWG interim recommendations to explore the coupling structure of
  a Higgs-like particle}},  \href{https://arxiv.org/abs/1209.0040}{{\ttfamily
  1209.0040}}.

\bibitem{LHCHiggsCrossSectionWorkingGroup:2013rie}
{\scshape LHC Higgs Cross Section Working Group} collaboration, J.~R. Andersen
  et~al., \emph{{Handbook of LHC Higgs Cross Sections: 3. Higgs Properties}},
  \href{https://arxiv.org/abs/1307.1347}{{\ttfamily 1307.1347}}.

\bibitem{deBlas:2018tjm}
J.~de~Blas, O.~Eberhardt and C.~Krause, \emph{{Current and Future Constraints
  on Higgs Couplings in the Nonlinear Effective Theory}},
  \href{https://doi.org/10.1007/JHEP07(2018)048}{\emph{JHEP} {\bfseries 07}
  (2018) 048} [\href{https://arxiv.org/abs/1803.00939}{{\ttfamily
  1803.00939}}].

\bibitem{Ciccolini:2007ec}
M.~Ciccolini, A.~Denner and S.~Dittmaier, \emph{{Electroweak and QCD
  corrections to Higgs production via vector-boson fusion at the LHC}},
  \href{https://doi.org/10.1103/PhysRevD.77.013002}{\emph{Phys. Rev. D}
  {\bfseries 77} (2008) 013002}
  [\href{https://arxiv.org/abs/0710.4749}{{\ttfamily 0710.4749}}].

\bibitem{Bierlich:2019rhm}
C.~Bierlich et~al., \emph{{Robust Independent Validation of Experiment and
  Theory: Rivet version 3}},
  \href{https://doi.org/10.21468/SciPostPhys.8.2.026}{\emph{SciPost Phys.}
  {\bfseries 8} (2020) 026} [\href{https://arxiv.org/abs/1912.05451}{{\ttfamily
  1912.05451}}].

\bibitem{Bierlich:2024vqo}
C.~Bierlich, A.~Buckley, J.~M. Butterworth, C.~Gutschow, L.~Lonnblad,
  T.~Procter et~al., \emph{{Robust independent validation of experiment and
  theory: Rivet version 4 release note}},
  \href{https://doi.org/10.21468/SciPostPhysCodeb.36}{\emph{SciPost Phys.
  Codeb.} {\bfseries 36} (2024) 1}
  [\href{https://arxiv.org/abs/2404.15984}{{\ttfamily 2404.15984}}].

\bibitem{Ohl:1998jn}
T.~Ohl, \emph{{Vegas revisited: Adaptive Monte Carlo integration beyond
  factorization}},
  \href{https://doi.org/10.1016/S0010-4655(99)00209-X}{\emph{Comput. Phys.
  Commun.} {\bfseries 120} (1999) 13}
  [\href{https://arxiv.org/abs/hep-ph/9806432}{{\ttfamily hep-ph/9806432}}].

\bibitem{Brass:2018xbv}
S.~Brass, W.~Kilian and J.~Reuter, \emph{{Parallel Adaptive Monte Carlo
  Integration with the Event Generator WHIZARD}},
  \href{https://doi.org/10.1140/epjc/s10052-019-6840-2}{\emph{Eur. Phys. J. C}
  {\bfseries 79} (2019) 344}
  [\href{https://arxiv.org/abs/1811.09711}{{\ttfamily 1811.09711}}].

\bibitem{ChokoufeNejad:2016qux}
B.~Chokouf\'e~Nejad, W.~Kilian, J.~M. Lindert, S.~Pozzorini, J.~Reuter and
  C.~Weiss, \emph{{NLO QCD predictions for off-shell $ t\overline{t} $ and $
  t\overline{t}H $ production and decay at a linear collider}},
  \href{https://doi.org/10.1007/JHEP12(2016)075}{\emph{JHEP} {\bfseries 12}
  (2016) 075} [\href{https://arxiv.org/abs/1609.03390}{{\ttfamily
  1609.03390}}].

\bibitem{Stienemeier:2021cse}
P.~Stienemeier, S.~Bra\ss{}, P.~Bredt, W.~Kilian, N.~Kreher, T.~Ohl et~al.,
  \emph{{WHIZARD 3.0: Status and News}},  in \emph{{International Workshop on
  Future Linear Colliders}}, 4, 2021,
  \href{https://arxiv.org/abs/2104.11141}{{\ttfamily 2104.11141}}.

\bibitem{Frixione:1995ms}
S.~Frixione, Z.~Kunszt and A.~Signer, \emph{{Three jet cross-sections to
  next-to-leading order}},
  \href{https://doi.org/10.1016/0550-3213(96)00110-1}{\emph{Nucl. Phys. B}
  {\bfseries 467} (1996) 399}
  [\href{https://arxiv.org/abs/hep-ph/9512328}{{\ttfamily hep-ph/9512328}}].

\bibitem{Frixione:1997np}
S.~Frixione, \emph{{A General approach to jet cross-sections in QCD}},
  \href{https://doi.org/10.1016/S0550-3213(97)00574-9}{\emph{Nucl. Phys. B}
  {\bfseries 507} (1997) 295}
  [\href{https://arxiv.org/abs/hep-ph/9706545}{{\ttfamily hep-ph/9706545}}].

\bibitem{Nason:2004rx}
P.~Nason, \emph{{A New method for combining NLO QCD with shower Monte Carlo
  algorithms}},
  \href{https://doi.org/10.1088/1126-6708/2004/11/040}{\emph{JHEP} {\bfseries
  11} (2004) 040} [\href{https://arxiv.org/abs/hep-ph/0409146}{{\ttfamily
  hep-ph/0409146}}].

\bibitem{Stienemeier:2022wmy}
P.~Stienemeier, \emph{{Automation and Application of fixed-order and matched
  NLO Simulations}}, Ph.D. thesis, Hamburg U., Hamburg, 2022.
\newblock 10.3204/PUBDB-2022-07425.

\bibitem{Nogueira:1991ex}
P.~Nogueira, \emph{{Automatic Feynman Graph Generation}},
  \href{https://doi.org/10.1006/jcph.1993.1074}{\emph{J. Comput. Phys.}
  {\bfseries 105} (1993) 279}.

\bibitem{Vermaseren:2000nd}
J.~A.~M. Vermaseren, \emph{{New features of FORM}},
  \href{https://arxiv.org/abs/math-ph/0010025}{{\ttfamily math-ph/0010025}}.

\bibitem{Kuipers:2012rf}
J.~Kuipers, T.~Ueda, J.~A.~M. Vermaseren and J.~Vollinga, \emph{{FORM version
  4.0}}, \href{https://doi.org/10.1016/j.cpc.2012.12.028}{\emph{Comput. Phys.
  Commun.} {\bfseries 184} (2013) 1453}
  [\href{https://arxiv.org/abs/1203.6543}{{\ttfamily 1203.6543}}].

\bibitem{vanDeurzen:2013saa}
H.~van Deurzen, G.~Luisoni, P.~Mastrolia, E.~Mirabella, G.~Ossola and
  T.~Peraro, \emph{{Multi-leg One-loop Massive Amplitudes from Integrand
  Reduction via Laurent Expansion}},
  \href{https://doi.org/10.1007/JHEP03(2014)115}{\emph{JHEP} {\bfseries 03}
  (2014) 115} [\href{https://arxiv.org/abs/1312.6678}{{\ttfamily 1312.6678}}].

\bibitem{Peraro:2014cba}
T.~Peraro, \emph{{Ninja: Automated Integrand Reduction via Laurent Expansion
  for One-Loop Amplitudes}},
  \href{https://doi.org/10.1016/j.cpc.2014.06.017}{\emph{Comput. Phys. Commun.}
  {\bfseries 185} (2014) 2771}
  [\href{https://arxiv.org/abs/1403.1229}{{\ttfamily 1403.1229}}].

\bibitem{vanHameren:2010cp}
A.~van Hameren, \emph{{OneLOop: For the evaluation of one-loop scalar
  functions}}, \href{https://doi.org/10.1016/j.cpc.2011.06.011}{\emph{Comput.
  Phys. Commun.} {\bfseries 182} (2011) 2427}
  [\href{https://arxiv.org/abs/1007.4716}{{\ttfamily 1007.4716}}].

\bibitem{Maltoni:2016yxb}
F.~Maltoni, E.~Vryonidou and C.~Zhang, \emph{{Higgs production in association
  with a top-antitop pair in the Standard Model Effective Field Theory at NLO
  in QCD}}, \href{https://doi.org/10.1007/JHEP10(2016)123}{\emph{JHEP}
  {\bfseries 10} (2016) 123}
  [\href{https://arxiv.org/abs/1607.05330}{{\ttfamily 1607.05330}}].

\bibitem{masterthesisMarijn}
M.~van Geest, ``{Anomalous Couplings within Standard Model Effective Field
  Theory in ttH Production at NLO QCD}.'' Master thesis, Karlsruhe Institute of
  Technology, https://www.itp.kit.edu/publications/diploma.

\bibitem{Cascioli:2011va}
F.~Cascioli, P.~Maierhofer and S.~Pozzorini, \emph{{Scattering Amplitudes with
  Open Loops}},
  \href{https://doi.org/10.1103/PhysRevLett.108.111601}{\emph{Phys. Rev. Lett.}
  {\bfseries 108} (2012) 111601}
  [\href{https://arxiv.org/abs/1111.5206}{{\ttfamily 1111.5206}}].

\bibitem{Buccioni:2019sur}
F.~Buccioni, J.-N. Lang, J.~M. Lindert, P.~Maierh\"ofer, S.~Pozzorini, H.~Zhang
  et~al., \emph{{OpenLoops 2}},
  \href{https://doi.org/10.1140/epjc/s10052-019-7306-2}{\emph{Eur. Phys. J. C}
  {\bfseries 79} (2019) 866}
  [\href{https://arxiv.org/abs/1907.13071}{{\ttfamily 1907.13071}}].

\bibitem{masterthesisJens}
J.~Braun, ``{Higgs Boson Pair Production in Vector Boson Fusion at NLO QCD in
  HEFT}.'' Master thesis, Karlsruhe Institute of Technology,
  https://www.itp.kit.edu/publications/diploma.

\bibitem{PDF4LHCWorkingGroup:2022cjn}
{\scshape PDF4LHC Working Group} collaboration, R.~D. Ball et~al., \emph{{The
  PDF4LHC21 combination of global PDF fits for the LHC Run III}},
  \href{https://doi.org/10.1088/1361-6471/ac7216}{\emph{J. Phys. G} {\bfseries
  49} (2022) 080501} [\href{https://arxiv.org/abs/2203.05506}{{\ttfamily
  2203.05506}}].

\bibitem{Buckley:2014ana}
A.~Buckley, J.~Ferrando, S.~Lloyd, K.~Nordstr\"om, B.~Page, M.~R\"ufenacht
  et~al., \emph{{LHAPDF6: parton density access in the LHC precision era}},
  \href{https://doi.org/10.1140/epjc/s10052-015-3318-8}{\emph{Eur. Phys. J. C}
  {\bfseries 75} (2015) 132} [\href{https://arxiv.org/abs/1412.7420}{{\ttfamily
  1412.7420}}].

\bibitem{Cacciari:2008gp}
M.~Cacciari, G.~P. Salam and G.~Soyez, \emph{{The anti-$k_t$ jet clustering
  algorithm}}, \href{https://doi.org/10.1088/1126-6708/2008/04/063}{\emph{JHEP}
  {\bfseries 04} (2008) 063} [\href{https://arxiv.org/abs/0802.1189}{{\ttfamily
  0802.1189}}].

\bibitem{Cacciari:2011ma}
M.~Cacciari, G.~P. Salam and G.~Soyez, \emph{{FastJet User Manual}},
  \href{https://doi.org/10.1140/epjc/s10052-012-1896-2}{\emph{Eur. Phys. J. C}
  {\bfseries 72} (2012) 1896}
  [\href{https://arxiv.org/abs/1111.6097}{{\ttfamily 1111.6097}}].

\bibitem{iminuit}
H.~Dembinski and P.~O. et~al., \emph{scikit-hep/iminuit}, .

\bibitem{James:1975dr}
F.~James and M.~Roos, \emph{{Minuit: A System for Function Minimization and
  Analysis of the Parameter Errors and Correlations}},
  \href{https://doi.org/10.1016/0010-4655(75)90039-9}{\emph{Comput. Phys.
  Commun.} {\bfseries 10} (1975) 343}.

\bibitem{LHCHWG24}
A.~R. Martinez, ``{Talk at the 21st Workshop of the {LHC} Higgs Working
  Group}.'' \url{https://indico.cern.ch/event/1389221/contributions/6195709/}.

\bibitem{Cohen:2021ucp}
T.~Cohen, N.~Craig, X.~Lu and D.~Sutherland, \emph{{Unitarity violation and the
  geometry of Higgs EFTs}},
  \href{https://doi.org/10.1007/JHEP12(2021)003}{\emph{JHEP} {\bfseries 12}
  (2021) 003} [\href{https://arxiv.org/abs/2108.03240}{{\ttfamily
  2108.03240}}].

\bibitem{Barducci:2023lqx}
D.~Barducci, M.~Nardecchia and C.~Toni, \emph{{Perturbative unitarity
  constraints on generic vector interactions}},
  \href{https://doi.org/10.1007/JHEP09(2023)134}{\emph{JHEP} {\bfseries 09}
  (2023) 134} [\href{https://arxiv.org/abs/2306.11533}{{\ttfamily
  2306.11533}}].

\bibitem{Jager:2025isz}
B.~J\"ager, A.~Karlberg and S.~Reinhardt, \emph{{Precision tools for the
  simulation of double-Higgs production via vector-boson fusion}},
  \href{https://arxiv.org/abs/2502.09112}{{\ttfamily 2502.09112}}.

\end{thebibliography}\endgroup

\end{document}